\documentclass[%
reprint,
superscriptaddress,
amsmath,amssymb,
aps,
prl,
]{revtex4-2}
\usepackage{ragged2e}
\usepackage[caption=false]{subfig}
\DeclareCaptionOption{justified}[]{\caption@setjustification{justified}}
\usepackage{titlesec}
\usepackage{xcolor}
\definecolor{darkblue}{rgb}{0, 0.08, 0.45}
\PassOptionsToPackage{colorlinks,allcolors=darkblue}{hyperref}
\usepackage{hyperref}   
\usepackage{orcidlink}  

\usepackage{graphicx}
\usepackage{dcolumn}
\usepackage{bm}
\usepackage[capitalise]{cleveref}

\usepackage{physics}
\usepackage[printwatermark]{xwatermark}
\usepackage{comment}
\usepackage{mathtools}
\usepackage{bbold}
\usepackage{soul}

\newcommand{\SU}{\mathrm{SU}}
\newcommand{\dg}{^{\dagger}}

\newif\ifSM
\SMtrue
\newif\ifmainText
\mainTexttrue 

\begin{document}
\setlength{\intextsep}{10pt plus 2pt minus 2pt}

\ifmainText
\title{\textbf{Antiferromagnetism and Stripe Channel Order in the $\mathrm{SU}(N)$-Symmetric \\ Two-Channel Kondo Lattice Model}}
 
\author{Elyasaf Y. Cohen \orcidlink{0000-0003-0441-2056}}
    \email{elyasaf.cohen1@mail.huji.ac.il}
    \affiliation{Racah Institute of Physics, The Hebrew University of Jerusalem, Jerusalem 91904, Israel}
\author{Fakher F. Assaad \orcidlink{0000-0002-3302-9243} }
    \affiliation{Institut für Theoretische Physik und Astrophysik, Universität Würzburg, 97074 Würzburg, Germany}
    \affiliation{Würzburg-Dresden Cluster of Excellence ctd.qmat, Am Hubland, 97074 Würzburg, Germany}
\author{Snir Gazit \orcidlink{0000-0002-2375-0440}}
    \email{snir.gazit@mail.huji.ac.il}
    \affiliation{Racah Institute of Physics, The Hebrew University of Jerusalem, Jerusalem 91904, Israel}
    \affiliation{The Fritz Haber Research Center for Molecular Dynamics, The Hebrew University of Jerusalem, Jerusalem 91904, Israel}

\date{\today}
\begin{abstract}
We carry out large-scale, sign-problem-free determinant quantum Monte Carlo simulations of the square lattice $\mathrm{SU}(N)$-symmetric two-channel Kondo lattice model at half-filling. We map out the zero-temperature phase diagram for $N = 2, 4, 6$, and $8$, as a function of the Kondo coupling strength. In the weak-coupling regime, we observe antiferromagnetic order of the localized moments. Remarkably, for $N \geq 6$, sufficiently strong Kondo coupling induces spontaneous channel symmetry breaking, forming a stripe dimerization pattern with a wave vector $\boldsymbol{k}=(\pi,0)$ alternating between channels. These findings are supported by a complementary large-$N$ saddle point analysis, which identifies the striped hybridization pattern as the energetically preferred configuration. The spatial symmetry breaking results in an anisotropic Fermi surface reconstruction. \\
\\
\noindent DOI: \href{https://doi.org/10.1103/n715-5r2y}{10.1103/n715-5r2y}
\end{abstract}
\maketitle

{\it Introduction---}The Kondo lattice model (KLM) is a paradigmatic model of strongly correlated fermions, capturing the interplay between Kondo-coupled localized moments and itinerant electrons. It is typically invoked as an effective model for heavy-fermion materials~\cite{Stewart_1984, Coleman_2007}. The resulting phase diagram displays particularly rich physical phenomena, including superconductivity, magnetic orders, unconventional criticality, and fractionalized Fermi liquids~\cite{Gegenwart_2008, Stewart_2017, Si_2010_book, Si_2010_Science, Si_2014, Senthil_2003, Coleman_1983, Doniach1977, Mazza24, Danu20, Danu22}. Curiously, the KLM has more recently been proposed as an effective model for strong correlation in magic-angle twisted bilayer graphene~\cite{Song_2022, Hu_2023}.

The two-channel Kondo lattice model (2CKLM) is a natural extension of the single-band KLM, where localized moments interact with two bands of itinerant electrons. The channel index introduces a new degree of freedom. Strong correlations can potentially lead to channel-selective hybridization, breaking channel symmetry~\cite{Hoshino_2011, Zhang_2018}. From the experimental perspective, the 2CKLM has been proposed as an effective description for UBe$_{13}$, URu$_2$Si$_2$, and 1-2-20 Pr-based materials~\cite{Cox_1987_uranium, Cox_1996_resistivity, Cox_1996, Sakai_2011, Tokunaga_2013, Yoshida_2017, Iwasa_2017, Zhang_2018}. 

\begin{figure}
    \captionsetup[subfigure]{labelformat=empty}
    \subfloat[\label{subfig:model}]{}
    \subfloat[\label{subfig:phase_diagram}]{}
    \centering
    \includegraphics[width=0.98\linewidth]{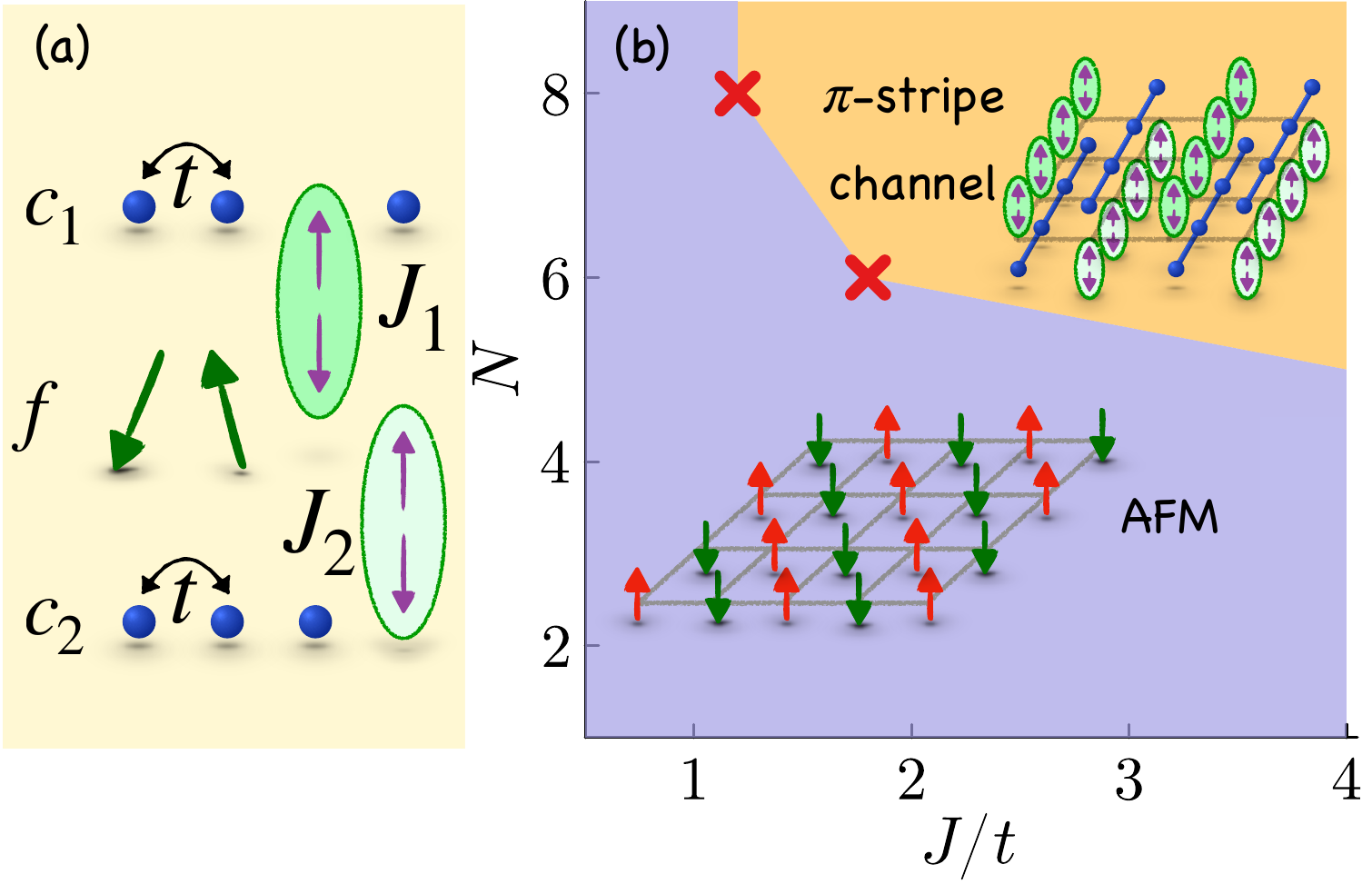}
    \vspace{-1em}
    \caption{(a) Schematic illustration of the 2CKLM,~\cref{eq:H_2CKLM}. Each site contains two itinerant electrons ($c_a$ with $a=1,2$) with intrachannel hopping of strength $t$ and a single localized moment ($f$) interacting via Kondo couplings $J_a$. (b) Quantum phase diagram of the $\SU(N)$ symmetric 2CKLM on the half-filled square lattice. At low $J/t$, we find AFM order (purple) mediated by RKKY interactions. At strong Kondo couplings $J/t\gg1$ and for $N\geq6$, we identify a channel-symmetry-breaking stripe phase (yellow) with an ordering wave vector $\bm{k}=(\pi,0)$.}
    \label{fig:phase_diag_and_model}
    \vspace{-2em}
\end{figure}

\begin{figure*}[!t]
    \captionsetup[subfigure]{labelformat=empty}
    \subfloat[\label{subfig:spin}]{}
    \subfloat[\label{subfig:dimer}]{}
    \subfloat[\label{subfig:imbalance}]{}
    \centering
        \includegraphics[width=0.32\textwidth]{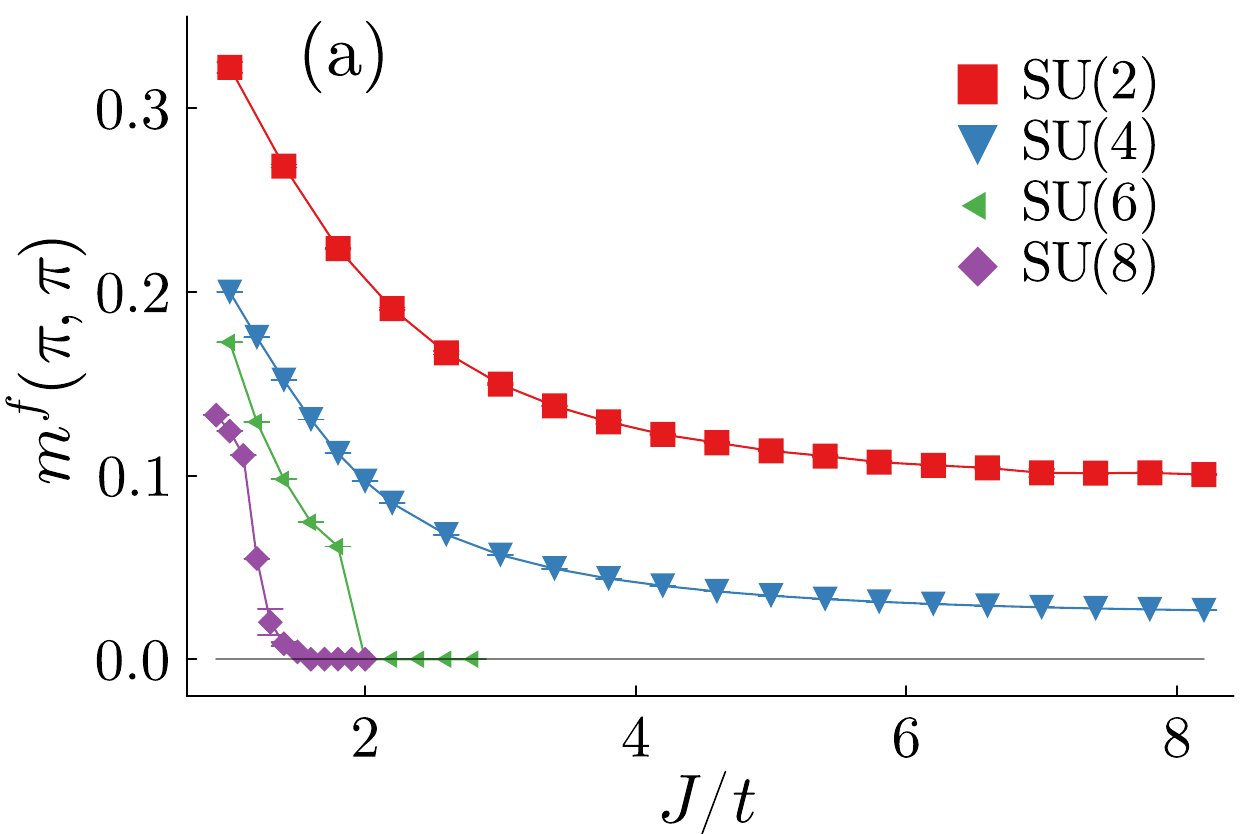}
        \includegraphics[width=0.32\textwidth]{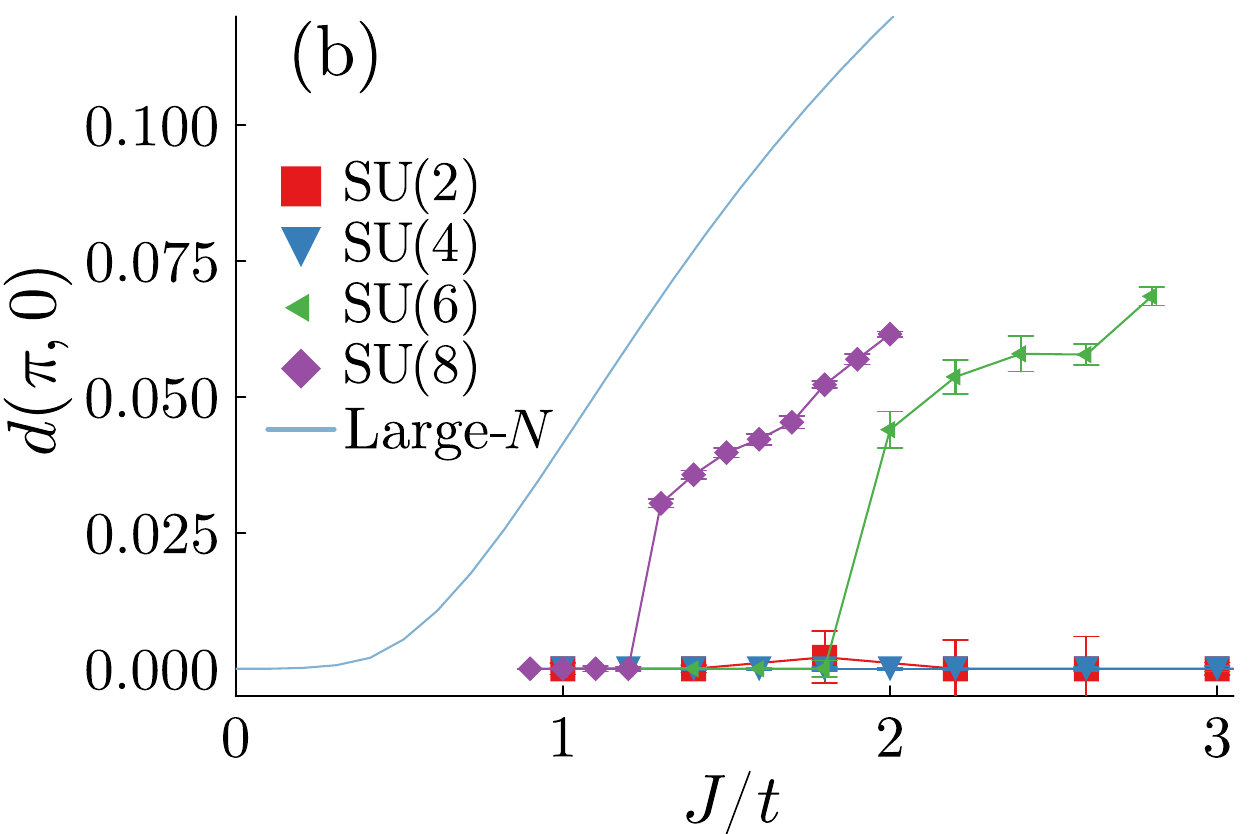}
        \includegraphics[width=0.32\textwidth]{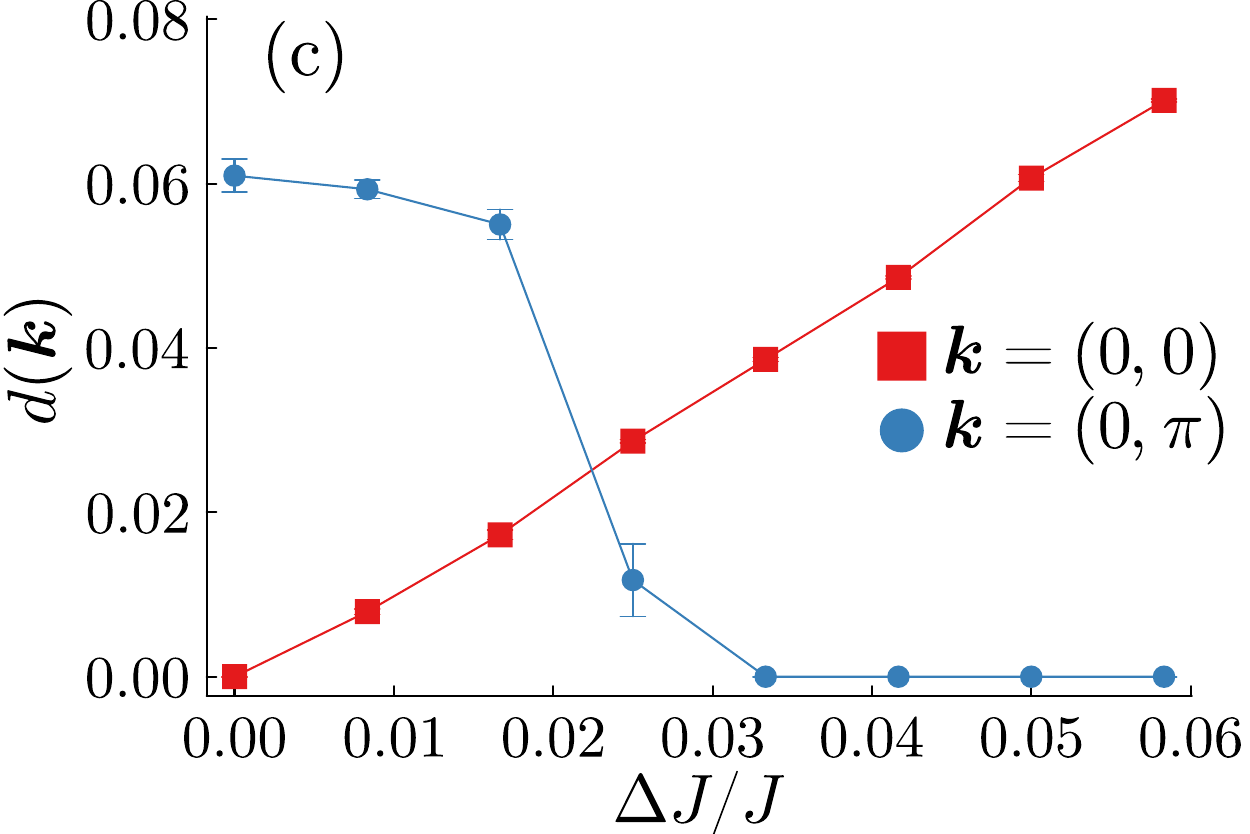}
        \vspace{-1em}
        \caption{(a) Staggered magnetization order parameter of localized moments. (b) Dimer order parameter at momentum $\bm{k}=(\pi,0)$. The blue line represents the dimer order parameter limit for $N\to\infty$, as obtained from the large-$N$ analysis [see End Matter]. In both (a) and (b), the x axis is the Kondo coupling $J$ in units of the hopping amplitude $t$. Different curves in (a) and (b) correspond to different values of $N$. (c) Dimer order parameter for different momenta $\bm{k}$ for $N=6$ at $J_1/t=2.4$, while $J_2=J_1-\Delta J$. All presented values result from an extrapolation to the thermodynamic limit, see~\cite{SM}.}
    \label{fig:order_parameters}
    \vspace{-2em}
\end{figure*}

The 2CKLM has been extensively studied theoretically in the large-$N$ limit for both $\SU(N)$ and $\mathrm{Sp}(N)$ symmetric models~\cite{Wugalter_2020, Flint_2008, Hoshino_2014, Assa_1986, Read_1983}, and numerically using the density matrix renormalization group (DMRG) in one dimension~\cite{Moreno_2001, Kornjaca_2024, Schauerte_2005} and the dynamical mean-field theory in the infinite-dimensional limit~\cite{Nourafkan_2008, Hoshino_2011, Koji_2020, Yang_2022}. The predicted low-temperature phase diagram includes channel symmetry breaking~\cite{Nourafkan_2008, Koji_2020, Yang_2022, Schauerte_2005}, unconventional superconductivity~\cite{Ott_1983, Hoshino_2014}, and a topological Kondo insulator~\cite{Hoshino_2012, Zhang_2018}. However, resolving the phase diagram using numerically exact techniques in dimensions greater than 1 remains an outstanding problem.

In this Letter, we study the low-temperature phase diagram of the half-filled $\SU(N)$ symmetric 2CKLM defined on a square lattice, for $N = 2,4,6,$ and $8$. To that end, we employ sign-problem-free determinant quantum Monte Carlo (DQMC) simulations, allowing us to obtain unbiased and numerically exact results~\cite{Assaad99a, Raczkowski_2020}. For weak Kondo couplings, we find an antiferromagnetic (AFM) order for all $N$ values, which persists even at relatively large Kondo couplings for $ N=2$ and $ N=4$. Remarkably, for $N\geq 6$ beyond a critical Kondo coupling, we identify channel symmetry breaking with a hybridization characterized by a $\pi$-modulated stripe pattern, as opposed to the more standard ferro-channel hybridization phase considered previously, see \cref{subfig:phase_diagram},~\cite{Wugalter_2020}. Analytical results in the large-$N$ limit corroborate the numerical findings of a stripe channel-symmetry-breaking ground state and the resulting reconstructed Fermi surface. 

{\it The model---}As a concrete microscopic model, see \cref{subfig:model}, we consider the $\SU(N)$-symmetric 2CKLM, defined on the square lattice,
\begin{align}
    \label{eq:H_2CKLM}
    H = -t \sum_{a=1}^2 \sum_{\left\langle \bm{i},\bm{j} \right\rangle,\mu}&  \qty(c^\dagger_{\bm{i}\mu a} c_{\bm{j}\mu a} +\text{H.c.}) \\
    &+ \sum_{a=1}^2 \frac{J_a}{N} \sum_{\bm{i}}\sum_{\mu \nu} 
    c^\dagger_{\bm{i}\mu a} c_{\bm{i}\nu a}
     {S}^{\bm{i}}_{\nu\mu}\,. \nonumber
\end{align}
Here, $c^\dagger_{\bm{i}\mu a}$($c_{\bm{i}\mu a}$) creates (annihilates) an electron at site $\bm{i}$ and channel $a=1,2$, carrying an $\SU(N)$ flavor $\mu=1,\ldots,N$. The first term describes nearest-neighbor hopping of itinerant electrons with amplitude $t$. Interchannel hopping is forbidden by construction. The second term represents exchange interactions of strength $J_a$ between itinerant electrons at channel $a$ and localized spins. Unless stated otherwise, we focus on the channel symmetric case $J_1=J_2=J$. The operators $S^{\bm{i}}_{\mu\nu}$ are a representation of generators of the $\SU(N)$ group. We tune to half-filling, which corresponds to a vanishing chemical potential. In addition to $\SU(N)$ spin symmetry, our model respects the $\mathrm{U}(1)$ symmetry corresponding to electron number conservation and the $\SU(2)$ symmetry associated with channel rotations~\cite{Wugalter_2020}.

In the weak coupling limit $J/t \ll 1$, the itinerant electrons induce an effective Ruderman–Kittel–Kasuya–Yosida (RKKY) interaction~\cite{Ruderman_1954, Coleman_1983} between the localized moments $S^{\bm{i}}_{\mu\nu}$, which at half-filling is purely antiferromagnetic, resulting in an AFM phase~\cite{Jarrell_cox_1997, Coleman_1983, Hoshino_2011}. By contrast, in the opposite limit, $J/t \gg 1$, previous large-$N$ studies~\cite{Wugalter_2020} predicted that the strong Kondo coupling leads to a selective formation of singlets between localized spins and itinerant electrons $c_a$ belonging to one of the channels. This results in a half-insulator half-metal state, which, crucially, breaks the channel symmetry. Away from half-filling, a spatial modulation of the channel binding was suggested, especially in the form of a staggered antiferrochannel configuration~\cite{Wugalter_2020, Kornjaca_2024, Zhang_2018}.

{\it Numerical simulations---}To numerically solve the 2CKLM in~\cref{eq:H_2CKLM}, we consider the fully antisymmetric fermionic representation of the $\SU(N)$ generators, constrained to half-filling,
\begin{equation}
    S^{\bm{i}}_{\mu\nu} = f_{\bm{i}\mu}^{\dagger} 
    f_{\bm{i}\nu} 
    - \frac{n^f_{\bm{i}}}{N}
    \delta_{\mu\nu}, 
    \quad 
    n^f_{\bm{i}} = \sum_{\mu}
    f^\dagger_{\bm{i}\mu} 
    f_{\bm{i}\mu} = N/2.
    \label{eq:constraint}
\end{equation}
In terms of the Abrikosov fermions $f_{\bm{i}\mu}$, the Hamiltonian reads,
\begin{align}
\label{eq:H_after_abrikosov}
    H = &-t \sum_{a=1}^2 \sum_{\left\langle \bm{i},\bm{j} \right\rangle,\mu}
    \qty(c^\dagger_{\bm{i}\mu a} c_{\bm{j}\mu a} +\text{H.c.}) \\
     &- \frac{J}{2N}\sum_{\bm{i}a\mu\nu} 
     \qty(c^\dagger_{\bm{i}\mu a} f_{\bm{i}\mu})
     \qty(f^\dagger_{\bm{i}\nu} c_{\bm{i}\nu a}) \nonumber
     +   
     \qty(f^\dagger_{\bm{i}\nu} c_{\bm{i}\nu a}) 
     \qty(c^\dagger_{\bm{i}\mu a} f_{\bm{i}\mu})
     \\
    &+ \frac{U_f}{N} \sum_{\bm{i}} \qty(n^f_{\bm{i}} - \frac{N}{2})^2. \nonumber
\end{align}
We note that a sufficiently strong Hubbard interaction, $U_f$, effectively enforces the half-filling constraint in~\cref{eq:constraint}~\cite{Raczkowski_2020}. 

For our solution, we employ the numerically exact DQMC technique~\cite{Gubernatis_Kawashima_Werner_2016}. Crucially, at half-filling~\cite{Raczkowski_2020} the model is free of the numerical sign problem for even values of $N$, allowing for unbiased simulations. We consider $N=2,4,6$, and $8$ flavors and linear system sizes up to $L=18$. To establish the zero temperature limit, we observed the convergence of observables by examining a series of decreasing temperatures. In practice, we found that $t \beta\geq4L$ is typically sufficient for all $J/t$ ratios. Lastly, we verified that our choice for Trotter step $\varepsilon$ is sufficiently small to ensure convergence within the error bars. In this work, we set $U_f/t=4$, and we verified that this ensures that the half-filling constraint in~\cref{eq:constraint} holds. The simulations were carried out using the ALF library~\cite{ALF}. Additional technical details and benchmarking are presented in~\cite{SM}. 

{\it Observables---}With the above discussion in mind, we consider the following physical observables to discriminate between the different phases. We probe the appearance of AFM order via the spin-spin correlation function of the localized moments, 
\begin{equation}
    S^{f}(\bm{k},L) = \frac{1}{L^4} 
    \sum_{\bm{i}\bm{j}\mu\nu} e^{-i\bm{k} \cdot ({\bm{i}} - {\bm{j}})} 
    \langle 
    S^{\bm{i}}_{\mu\nu}
    S^{\bm{j}}_{\nu\mu}
    \rangle,
    \label{eq:ss_corr}
\end{equation}
where the indices $\mu,\nu$ iterate over the $\SU(N)$ flavors. Focusing on a staggered configuration, we fix $\bm{k}=(\pi,\pi)$. 
The associated order parameter is the average magnetization, given by $m^f(\bm{k}) = \sqrt{\lim_{L \to \infty} S^{f}(\bm{k}, L)/(N^2-1)}$.

Anticipating channel symmetry breaking, we consider the channel magnetization, given by,
\begin{equation}
    \label{eq:M_vec}
    \vec{M}_{\bm{i}} = \langle 
    c_{\bm{i}\mu a}^\dagger 
    \vec{\tau}_{aa'} 
    c_{\bm{i}\mu'a'} 
    ({T}_{\mu\mu'}^\gamma 
    {T}_{\nu\nu'}^\gamma
    )
    f^\dagger_{\bm{i}\nu}
    f_{\bm{i}\nu'}
    \rangle.
\end{equation}
Here $\vec{\tau} = (\tau_x, \tau_y, \tau_z)$ are the standard Pauli matrices associated with the channel space, ${T}^\gamma$ are the $N^2-1$ generators of $\SU(N)$ with the normalization condition $\Tr[T^\gamma T^{\gamma'}]=\delta_{\gamma\gamma'}/2$. Repeated indices belonging to the flavor ($\mu,\nu$), channel ($a$), and generator ($\gamma$) are implicitly summed over. The $z$ component of $\vec{M}_{\bm{i}}$ measures the imbalance between singlet formation across channels. The associated dimer-dimer correlation function at momentum $\bm{k}$ is then:
\begin{equation} \label{eq:dimer}
    D(\bm{k},L) = \frac{1}{L^4} 
    \sum_{\bm{i},\bm{j}} 
    e^{-i{\bm{k}} \cdot ({\bm{i}} - {\bm{j}})} 
    \langle M_{\bm i}^z M_{\bm{j}}^z \rangle.
\end{equation}
As before, we define the channel-symmetry-breaking order parameter as $d(\bm{k}) = \sqrt{\lim_{L \to \infty} {D(\bm{k}, L)}/{{(N^2-1)^2}}}$. We note that for $k_x \neq k_y$, we define $D(\bm{k},L)$ as the sum of all configurations related by the square lattice ${C}_4$ symmetry transformations. 

{\it Numerical results---}We first compute AFM correlations of the localized moments ${m^f[\bm{k}=(\pi,\pi)]}$, and the channel order $d(\bm{k})$, for $N = 2, 4, 6$, and $8$. The presented results in~\cref{fig:order_parameters} are extrapolated to the thermodynamic limit $L\to\infty$, see~\cite{SM}.

Consistent with the above predictions, in the weak Kondo coupling limit $J/t \ll 1$, we find a finite AFM order for all $N$ values. Interestingly, for both $N=2$ and $N=4$, AFM correlations persist up to relatively large coupling strengths, reaching $J/t = 8.2$. In addition, for the range of coupling constants considered, we do not find evidence for long-range channel magnetization at any wave vector. These findings suggest that, for $N=2$ and $4$, AFM order persists for a wide range of couplings without evidence of channel magnetization.

\begin{figure}[t]
    \centering
    \includegraphics[width=\linewidth]{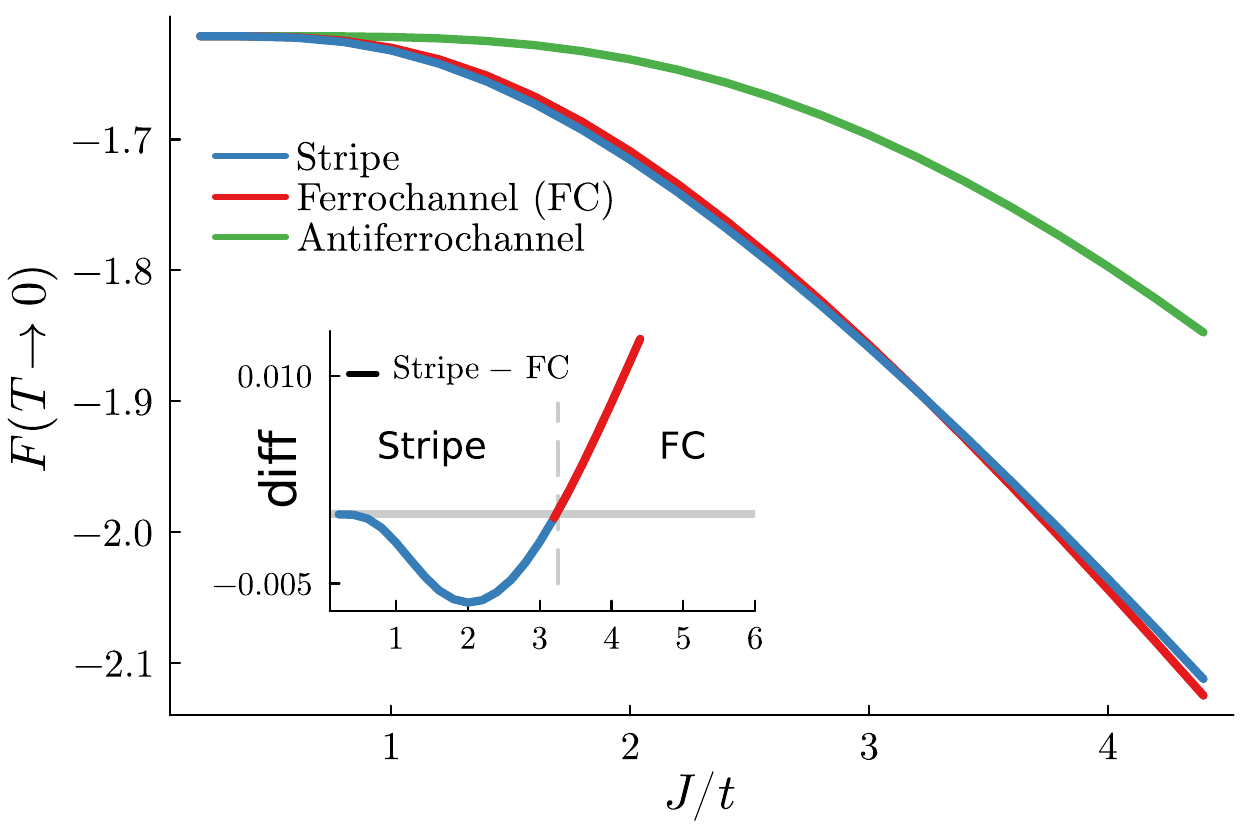}
    \caption{Ground-state energy per site and flavor, as a function of $J/t$, in the large-$N$ limit. Different curves correspond to different spatial \textit{Ans\"atze} of the hybridization field. The inset depicts the difference between the competing ferrochannel and $\pi$-modulated stripe channels.}
    \label{fig:free_energy_order}
\end{figure}

By contrast, for $N=6$ and $8$, AFM order vanishes at a critical coupling $J^{\textrm{AFM}}_c/t = 2.0(2)$ and $J^{\textrm{AFM}}_c/t = 1.6(1)$, respectively, as shown in~\cref{subfig:spin}. The vanishing of AFM order is expected to nucleate channel-symmetry-breaking order as the Kondo coupling increases. To see if this is indeed the case, we probe the dimer order parameter, $d(\bm{k})$, as a function of momenta. Unexpectedly, we observe channel-symmetry-breaking order at wave vectors $\bm{k}=(0,\pi)/(\pi,0)$, indicating a $\pi$-modulated stripe order phase, in contrast to the uniform or staggered patterns typically considered, see~\cite{SM}. Motivated by these results, in~\cref{subfig:dimer}, we depict the channel magnetization stripe order parameter, $d[\bm{k}=(\pi,0)]$ as a function of the Kondo coupling. From this analysis, we estimate the critical Kondo coupling for the onset of stripe order to be $J^{\textrm{stripe}}_c/t = 1.8(2)\approx J^{\textrm{AFM}}_c/t$ and $J^{\textrm{stripe}}_c/t = 1.3(1)<J^{\textrm{AFM}}_c/t$, for $N=6$ and $8$, respectively, see~\cref{subfig:dimer}. For $N=6$, the transition appears to be first order as reflected in the jump discontinuity of the relevant order parameters. Moreover, within our numerical resolution, we observe a coexistence of AFM and channel stripe order for $N=8$.

\begin{figure*}[t!]
    \centering
    \captionsetup[subfigure]{labelformat=empty} 
    \subfloat[\label{subfig:fermi_ferro}]{}
    \subfloat[\label{subfig:fermi_stripe}]{}
    \subfloat[\label{subfig:fermi_largeN}]{}
    \includegraphics[width=0.99\linewidth]{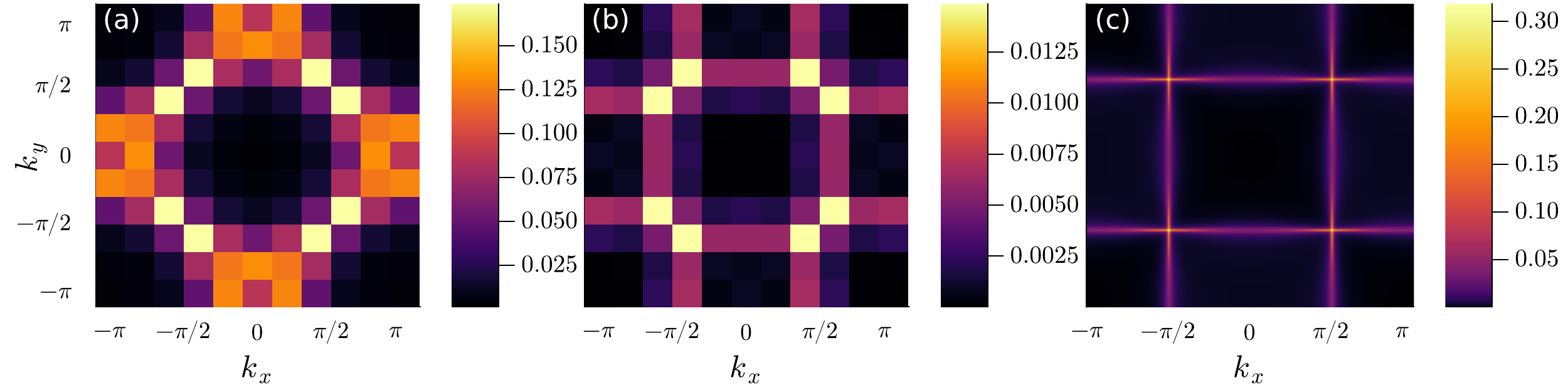}
    \vspace{-1em}
    \caption{Single-particle residue of $c$, evaluated via the imaginary-time Green function $G_c(k_x,k_y,\tau=\beta/2)$ for $L=10$ at (a) $J/t=1.4$ and (b) $J/t=2.4$ for $\SU(6)$. Simulations are carried out at inverse temperatures  $t\beta=40$. (c) $A(\bm{k},\omega=i0^+)$ evaluated in the large-$N$ limit for a $\pi$-modulated stripe channel \textit{Ansatz} and $J/t=3$. The two broken symmetry orientations are superimposed.}
    \label{fig:Green}
    \vspace{-2em}
\end{figure*}

To further probe the stability of the stripe channel order, we introduce explicit channel imbalance by setting $J_1=J$ and $ J_2=J-\Delta J$, which favors hybridization with the first channel. In~\cref{subfig:imbalance}, we consider $N=6$, and hybridization coupling $J=2.4>J_c$, where stripe channel correlations have already settled. We compare the uniform $d[\bm{k}=(0,0)]$ and stripe $d[\bm{k}=(\pi,0)]$ channel orders. We find that at a weak imbalance $\Delta J/J\ll 1$, stripe channel order persists and ferrochannel correlations weakly rise due to the explicit symmetry breaking. Only at a finite $\Delta J/J=0.025(5)$ stripe order is destroyed and replaced by a selective hybridization of the first channel. This result indicates the stability of the stripe channel phase in the presence of explicit channel asymmetry. 

Motivated by the above analysis, we seek to support our numerical findings of the stripe phase with an analytical calculation in the large-$N$ limit of~\cref{eq:H_after_abrikosov}~\cite{Coleman_2015_book, Wugalter_2020}. In our analysis, we compare three \textit{Ans\"atze} corresponding to different static hybridization patterns: ferrochannel [$\bm{k}=(0,0)$], antiferrochannel [$\bm{k}=(\pi,\pi)$], and stripes with channel modulating at wave vector $\bm{k}=(\pi,0)$. The results of this analysis are shown in~\cref{fig:free_energy_order}. Remarkably, we find that for low and intermediate Kondo coupling strength $J/t\leq3.23$, the stripe channel order has the lowest free energy, in agreement with our numerical results. However, further increasing the Kondo couplings results in the ground state favoring a uniform ferrochannel pattern, which is not observed in our numerical results. For comparison with the QMC results, in~\cref{subfig:dimer} we also plot the large-$N$ prediction for the channel symmetry breaking order parameter.

The large-$N$ analysis also allows us to compute the reconstructed band structure arising from channel symmetry breaking. To compare with DQMC results, we estimate the quasiparticle residue of the itinerant electrons by computing the imaginary-time single-particle Green's function $G_c(\bm{k}, \tau)=\left \langle c_a(\tau) c_a^\dagger \right \rangle$, evaluated at the maximal imaginary time difference $\tau=\beta/2$. The quantity $\beta G_c(\bm{k}, \tau=\beta/2)$ serves as a proxy for the single-particle residue, as it integrates over the spectral function across a small frequency window of order $T$ about the Fermi surface~\cite{Trivedi_1995, Berg_2019}. As a complementary analysis, we compute the spectral function in the large-$N$ limit defined as the imaginary part of the Green's function $A(\bm{k},\omega=i0^+) =-\frac{1}{\pi}\mathrm{Im}\left[ G_c(\bm{k},\omega=i0^+)\right],$ derived from the mean field Hamiltonian~[see End Matter~\cref{eq:H_mean_field}]. 

In~\cref{fig:Green}, we present this analysis for $N=6$. We observe that the Fermi surface evolves from a large, square-lattice-like Fermi surface in the AFM phase~\footnote{The AFM order in the $c$ channel sets in at much lower temperatures beyond our numerical resolution.}, see~\cref{subfig:fermi_ferro}, into a reconstructed Fermi surface that matches the one obtained from the stripe hybridization pattern in the large-$N$ analysis, see~\cref{subfig:fermi_stripe,subfig:fermi_largeN}. For the latter, we superimposed the contributions for the two broken symmetry orientations $\bm{k}=(0,\pi)$ and $\bm{k}=(\pi,0)$. The reconstructed Fermi surface indicates the emergence of one-dimensional metallic channels consistent with the stripe-channel order. The spectral weight along $\bm{k}_{x/y}=\pi/2$ matches the Fermi points crossing, expected at half-filling. These results should be contrasted with the predicted ferrochannel hybridization pattern, which is expected to produce a large Fermi surface corresponding to a half-metal half-insulator phase. 

{\it Summary and discussion---}In this work, we determined the ground-state phase diagram of the $\SU(N)$ symmetric 2CKLM defined on a two-dimensional square lattice at half-filling. We observed an AFM phase for weak Kondo couplings for all $N$ values. Remarkably, for $N\ge6$ and a sufficiently strong Kondo coupling, we identify a first order phase transition toward a channel-symmetry-broken phase, characterized by stripes modulated at wave vectors $(0,\pi)/(\pi,0)$. We complement our analysis via an analytic calculation in the large-$N$ limit, where we find a range of Kondo couplings for which hybridization with a stripe modulation minimizes the ground-state energy. In particular, this configuration is energetically favored over the formerly explored ferro-channel phase~\cite{Wugalter_2020}. Lastly, the reconstructed Fermi surface in the stripe channel, predicted by the large-$N$ analysis, is in agreement with our numerical results.

For the single-channel KLM, it has been shown that the QMC results for even values of $N$ smoothly map onto the mean-field result in the $\SU(N)$ symmetric Kondo phase~\cite{Raczkowski_2020}. In the 2CKLM, the Hamiltonian admits an $\SU(2)$ channel and $\SU(N)$ spin symmetries, and the large-$N$ saddle point does not take into account fluctuations transverse to the quantization axis of the channel symmetry breaking. These channel-wave Goldstone modes encode the meandering of stripes that host one-dimensional conduction channels. Studies of such smectic metals are a cornerstone in the description of the striped phase proposed in the realm of high $T_c$ superconductivity~\cite{Emery00}.

From an experimental standpoint, we showed that the reconstructed Fermi surface in the stripe phase yields a distinct signature in the spectral function $A_{c}(\bm{k},\omega)$, directly accessible via angle-resolved photoemission spectroscopy (ARPES). The emergence of quasi-one-dimensional conduction channels is manifested as Fermi surface segments parallel to the reciprocal axes. These features form a characteristic crossing pattern when horizontal and vertical symmetry-breaking domains are superimposed. This behavior is notably reminiscent of the spectral features observed via ARPES in the stripe-ordered cuprate (La$_{1.28}$Nd$_{0.6}$Sr$_{0.12}$)CuO$_4$~\cite{Zhou_1999}.

Experimental realizations of exact $\SU(N)$ symmetry for large $N$ include cold atomic systems~\cite{gorshkov2010two, Zhang_2014, scazza_2014}, multiorbital materials~\cite{Sasaki_2004, jarillo_2005}, and more recently, magic-angle twisted bilayer graphene~\cite{Chou_2023, Yantao_2024}. However, reaching sufficiently large $N$ values for which RKKY interactions and the resulting AFM instabilities are suppressed~\cite{Raczkowski_2020} in favor of channel symmetry breaking is challenging. Alternative physical mechanisms that suppress magnetic order, such as geometrical frustration, could offer a pathway to realizing the predicted stripe phase in experimentally relevant setups with small $N$ values.

From a complementary perspective, it is interesting to understand how to destabilize the observed striped phase. We have shown that a very small uniform channel field that reduces the $\SU(2)$ channel symmetry to $\mathrm U(1)$ suffices to destabilize the striped phase in favor of the uniform one. This close competition between various phases is also seen in our large-$N$ analysis. Taking into account $1/N$ corrections can better resolve this competition, in light of the preference for stripe ordering observed in our numerical results.

The channel-symmetry-breaking pattern involves fractionalized $f$ electrons, distinguished from the standard symmetry-breaking pattern known as ``order fractionalization'' as pointed out by recent theoretical analysis~\cite{Komijani_2019order, Wugalter_2020}. A key feature of this effect is the nonlocal behavior of the self-energy, which can be studied by analyzing the long-time behavior of the imaginary-time-dependent Green's function.

Because of the presence of the numerical sign problem, our simulations are restricted to half-filling. To test the stability of our results away from this limit at a finite chemical potential, one can invoke a large-$N$ analysis or numerical calculations using DMRG methods. Additionally, changing the lattice geometry to a honeycomb lattice would introduce Dirac cones in the band structure, potentially enriching our phase diagram~\cite{Shao_2024, Yang_2024}. We leave these interesting research directions to future studies.

\begin{acknowledgments}
{\it Acknowledgments}---We thank Assa Auerbach, Dror Orgad, and Maxim Khodas for fruitful discussions. This work is dedicated to the memory of our late colleague, Prof. Assa Auerbach. E.Y.C. acknowledges the support of the Council for Higher Education Scholarships Program for Outstanding Doctoral Students in Quantum Science and Technology, and a donor from the United Kingdom who has chosen to remain anonymous. S.G. acknowledges support from the Israel Science Foundation (ISF) Grant No. 586/22. Computational resources were provided by the Intel Labs Academic Compute Environment and the Fritz Haber Center for Molecular Dynamics, The Hebrew University of Jerusalem.  
F.F.A. acknowledges financial support from the German Research Foundation (DFG) under the Grant No. AS 120/16-1 (Project No. 493886309) that is part of the collaborative research project SFB QMS funded by the Austrian Science Fund (FWF) [Grant DOI: 10.55776/F86], as well as from the W\"urzburg-Dresden Cluster of Excellence {\it ctd.qmat} (EXC 2147, Project No.~390858490)
\end{acknowledgments}
{\it Data Availability---}The data used to generate the figures of this study are available at~\cite{zenodo_data}.

\makeatletter
\def\@bibsetup#1{\relax}
\makeatother
\bibliography{Kondo}

\newpage
\appendix
\onecolumngrid
\section{End Matter}
\twocolumngrid

\textit{Large-$N$ analysis---}In this section, we detail the large-$N$ analysis presented in the main text. We follow a standard approach, see, e.g.,~\cite{Coleman_2015_book}, starting from the coherent state path integral representation of the free energy associated with~\cref{eq:H_2CKLM} in the main text. To decouple the four-fermion interaction term, we perform a Hubbard-Stratonovich transformation. This involves introducing a complex auxiliary field $V_{\bm{i}a}$, yielding,
\begin{equation}
    Z = \int \mathcal{D}[\bar{c},c,\bar f,f,\bar V,V, \lambda] e^{-S},
\end{equation}
with the action,
\begin{align}
    S = &\int_0^\beta~d\tau  \nonumber
    \Bigg[
    \sum_{\bm{i} \mu a} 
    \bar c_{\bm{i} \mu a} \partial_\tau c_{\bm{i} \mu a} -
    t \sum_{\expval{\bm{i},\bm{j}} \mu a} 
    (\bar c_{\bm{i}\mu a} c_{\bm{j} \mu a} + \text{H.c.}) \\ 
    &+ \sum_{\bm{i} \mu a}
    \Big[ \bar c_{{\bm{i}}\mu a} \qty(V_{\bm{i}a} f_{\bm{i} \mu}) 
    + \qty(\bar V_{\bm{i}a} \bar f_{\bm{i}\mu}) c_{{\bm{i}}\mu a}\Big] \\
    &+ \sum_{\bm{i}\mu} \bar f_{\bm{i} \mu}
    \qty(\partial_\tau + \lambda_{\bm{i}}) 
    f_{\bm{i}\mu} 
    + \sum_{{\bm{i}}a} \frac NJ \abs{V_{\bm{i}a}}^2
    - \sum_{\bm{i}} \lambda_{\bm{i}} Q
    \Bigg]. \nonumber
\end{align} 
Here, $\mu$ is the $\SU(N)$ spin flavor index, the Lagrange multiplier $\lambda_{\bm{i}}$ enforces the on-site half-filling constraint $\sum_\mu f\dg_{\bm{i}\mu} f_{\bm{i}\mu} = N/2$, which implies $Q=N/2$. The auxiliary field $V_{\bm{i}a}$ acts as a dynamic hybridization between the conduction electrons $(c)$ and localized moments $(f)$. Because of the two-channel structure, the channel index of the hybridization parameter $\begin{pmatrix}
    V_{\bm{i},1}\ \ V_{\bm{i},2}
\end{pmatrix}$, up to its magnitude, behaves as a spinor.

\begin{figure*}
    \captionsetup[subfigure]{labelformat=empty}
    \subfloat[\label{subfig:bands_ferro}]{}
    \subfloat[\label{subfig:bands_stripe}]{}
    \subfloat[\label{subfig:min_v}]{}
    \centering
    \includegraphics[width=0.56\linewidth]{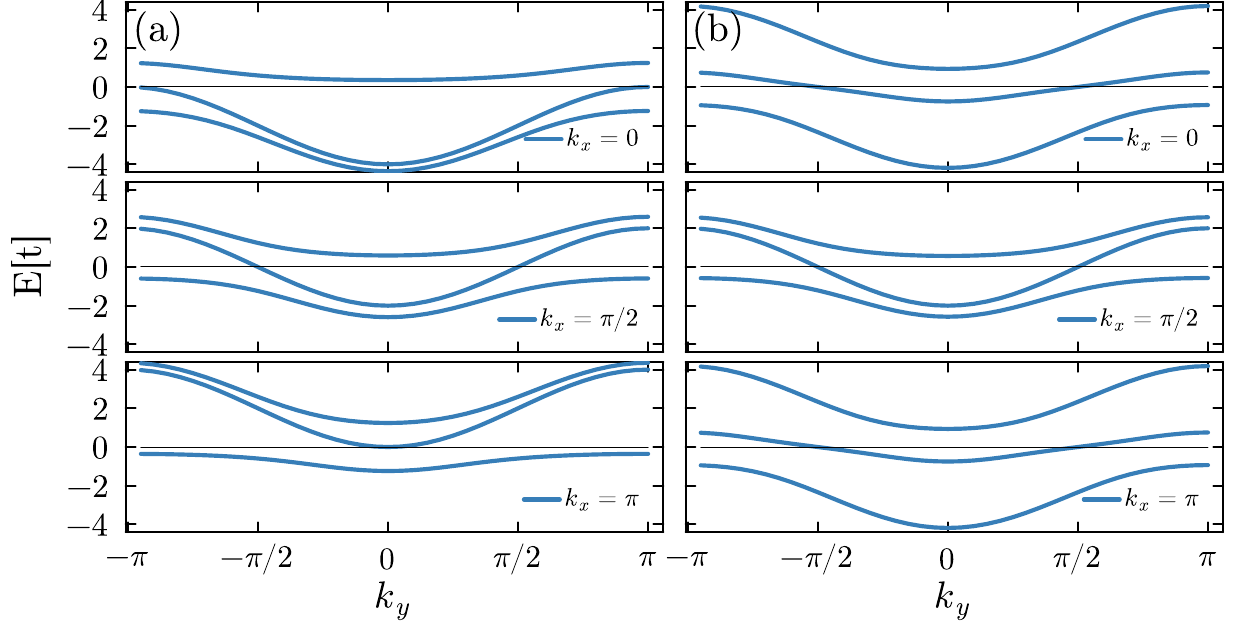}
    \includegraphics[width=0.43\linewidth]{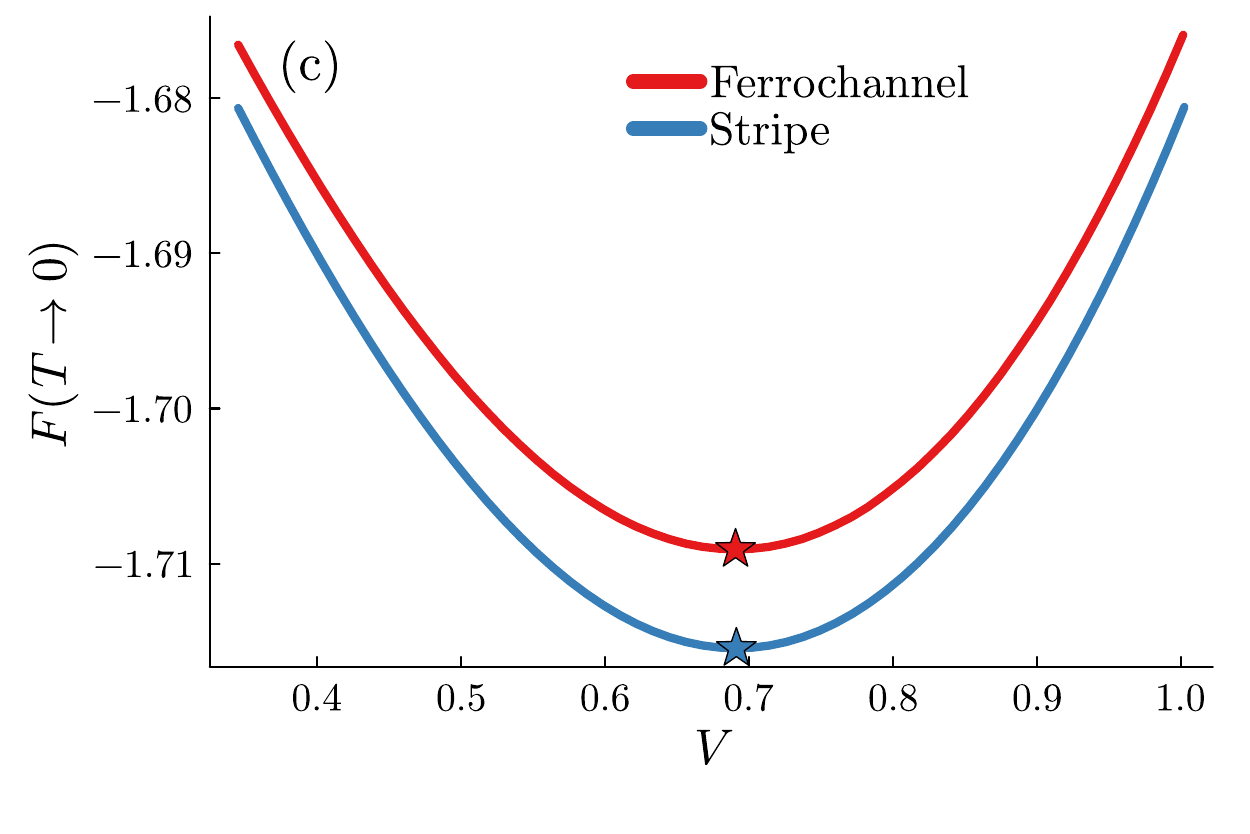}
     \caption{Band structure along several momentum cuts for $J/t=3$ obtained from the mean-field Hamiltonian in the large-$N$ limit. (a) ferrochannel and (b) $(\pi,0)$-modulated stripe channel \textit{Ans\"atze}. Energies are measured in units of~$t$. The chemical potential ($\mu=0$) is marked in a black line. We note that in the striped phase, the Fermi surface crossing is at momentum $\bm{k}=(k_x,\pi/2)$ for all $k_x$. (c) Zero-temperature free energy as a function of the hybridization field~$V$ for $J/t=2$, for which the free energy difference between the ferrochannel and stripe \textit{Ansatz} is maximal. The optimal hybridization parameter $V^*$ is marked by a star. }
    \label{fig:Bands_in_k} 
\end{figure*}

In the large-$N$ limit, we seek to determine the state that minimizes the free energy, and in the limit of $T\to0$, the ground-state energy. At half-filling, $\lambda_{\bm{i}}=0$, as can be seen from the saddle point equations. We further assume static hybridization fields, $V_{\bm{i}a}(\tau)=V_{\bm{i}a}$, with several spatial patterns. Motivated by our numerical simulations in the main text, we compare the ground-state energies of three specific spatial patterns: ferrochannel, antiferrochannel, and $\pi$-modulated stripe channel configurations. For each \textit{Ansatz}, we find the effective large-$N$ mean field Hamiltonian and solve for its band structure energy $E^b_{\bm{k}}[\{V\}_{\bm{i}a}]$.
The free energy per site and flavor is then given by,
\begin{equation}
    \frac{F}{L^2 N} = -T 
    \sum_{\bm{k},b\in\textrm{bands}} 
        \ln{\qty(1+e^{-\beta E^b_{\bm{k}}[\{V\}_{\bm{i}a}]})} 
    + \frac{1}{JL^2}\sum_{\bm{i}a}V_{\bm{i}a}^2.
\end{equation}
For each case, this procedure yields a free energy $F(V)$ where $V$ is the magnitude of the nonvanishing hybridization field. We then numerically minimize $F(V)$ with respect to $V$ to find the optimal value, $V^*$. As shown in~\cref{fig:free_energy_order} of the main text, we find that the stripe channel order attains the lowest free energy for low to intermediate coupling strengths $J/t\leq3.23$, whereas the ferrochannel phase becomes favorable at larger $J/t>3.23$. 

\textit{Ferrochannel Ansatz}---First, we consider the ferrochannel state, in which all $f$ electrons hybridize exclusively with a single conduction channel. Any spatially uniform hybridization, where $V_{\bm{i}a}$ is constant, is equivalent via a global $\SU(2)$ channel rotation to the simple ferrochannel state characterized by $V_{\bm{i},1}=V$ and $V_{\bm{i},2}=0$~\cite{Wugalter_2020}.
This \textit{Ansatz} yields a metallic band, along with two additional energy bands that closely resemble those found in the single-channel KLM~\cite{Coleman_2015_book}. The resulting band structure is given by,
\begin{equation} \label{eq:energy_band_FC}
    E^0_{\bm{k}} = \epsilon_{\bm{k}}, E^\pm_{\bm{k}} = \frac{\epsilon_{\bm{k}}}{2} \pm \sqrt{\qty(\frac{\epsilon_{\bm{k}}}{2})^2 + V^2},
\end{equation}
with the single-particle dispersion for the square lattice,
\begin{equation} \label{eq:2D_dispersion_realtion}
    \epsilon_{\bm{k}}= -2t \qty[\cos(k_x) + \cos(k_y)].
\end{equation}

\textit{$\pi$-modulated stripe channel Ansatz}---The second \textit{Ansatz} we consider is the $\pi$-modulated stripe channel state, where the hybridization alternates between the two channels along the $x/y$ direction. A simple representation for such a state, with stripes running along the $y$ direction, is
\begin{equation}
    V_{\bm{i}1}=\frac{V}{2}\qty(1+e^{i\bm{qi}}), \quad V_{\bm{i}2}=\frac{V}{2}\qty(1-e^{i\bm{qi}}),
\end{equation}
with $\bm{q}=(\pi,0)$. This choice sets a finite hybridization $V$ for the first (second) channel on even (odd) $i_x$, and otherwise vanishing.
The resulting action is then,
\begin{align}
    S &= \int_0^\beta~d\tau \Bigg[
     \sum_{\bm i a}\bar c_{\bm{i} a} \partial_\tau c_{\bm{i} a}
    + \sum_{\bm i}\bar f_{\bm{i} a} \partial_\tau f_{\bm{i} a}  \nonumber \\
    &- t \sum_{\expval{\bm{i},\bm{j}} a} (\bar c_{\bm{i}, a} c_{\bm{j}, a} + \text{h.c.})  \\
    &+ \frac{V}{2}
    \sum_{\bm{i}}
    \bigg( \bar c_{\bm{i},1} f_{\bm{i}} \qty(1+e^{i\bm{qi}})
    + \bar c_{\bm{i},2} f_{\bm{i}}\qty(1-e^{i\bm{qi}}) + \text{h.c.}\bigg) \nonumber
    \Bigg] \\ \nonumber
     & \quad \quad \quad + \beta N L^2 \frac{V^2}{J}.
\end{align}
The hybridization pattern leads to momentum scattering between momenta $\bm{k}$ and $\bm{k}+\bm{q}$. Consequently, in the momentum space representation, we sum over a reduced Brillouin zone $BZ/2=\bm{k}\in [0,\pi]\times[-\pi,\pi]$, giving rise to
\begin{equation}
    S = \int_0^\beta~d\tau 
    \sum_{\bm{k}\in BZ/2} \Psi^\dagger \qty[
    \partial_\tau + H_{\textrm{MF}}
    ] \Psi
    +\beta N L^2 \frac{V^2}{J},
\end{equation}
with 
\begin{equation}
\begin{split}
    \Psi=\begin{pmatrix}
        c_{\bm{k},1} \\
        c_{\bm{k},2} \\
        f_{\bm{k}} \\
        c_{\bm{k}+\bm{q},1} \\
        c_{\bm{k}+\bm{q},2} \\
        f_{\bm{k}+\bm{q}} 
    \end{pmatrix} ,H_{\textrm{MF}} = \begin{pmatrix}
        \epsilon_{\bm k}  &  0                  & \tilde{V}   & 0                         &  0                          & \tilde{V}  \\
        0                 &  \epsilon_{\bm k}   & \tilde{V}   & 0                         &  0                          & -\tilde{V}   \\
         \tilde{V}              &   \tilde{V}               & 0     & \tilde{V}                      &  -\tilde{V}                        & 0     \\
        0                 &  0                  & \tilde{V}  & \epsilon_{\bm k + \bm q}  &  0                          &  \tilde{V}  \\
        0                 &  0                  & -\tilde{V}   & 0                         &  \epsilon_{\bm k + \bm q}   &  \tilde{V}  \\
        \tilde{V}              &  -\tilde{V}                & 0     & \tilde{V}                      &   \tilde{V}                       &  0  
    \end{pmatrix},
\end{split}    \label{eq:H_mean_field}
\end{equation}
$\epsilon_{\bm k}$ as in~\cref{eq:2D_dispersion_realtion}, and $\tilde V=\frac{V}{2}$. The energy bands are obtained as the solution to a third-degree polynomial equation
\begin{align} \label{eq:energy_equation}
    &E^3 - 
    E^2 \qty({\epsilon_{\bm{k}}}+{\epsilon_{\bm{k}+\bm{q}}}) \\
    &+E\left({\epsilon_{\bm{k}}} {\epsilon_{\bm{k}+\bm{q}}} -V^2\right) +
    \frac{V^2}{2} \qty(
    {\epsilon_{\bm{k}}}+{\epsilon_{\bm{k}+\bm{q}}}
    ) = 0, \nonumber
\end{align} 
with twofold degeneracy for each band. In~\cref{subfig:bands_ferro,subfig:bands_stripe}, we show the band structure in the reduced Brillouin zone for both the ferrochannel and the stripe order.

The mean-field Hamiltonian is invariant under the particle-hole transformation 
\begin{equation}
    c\dg_{\bm{i}a} \to c_{\bm ia} (-1)^{i_x+i_y}, \quad
    f\dg_{\bm{i}a} \to -f_{\bm ia} (-1)^{i_x+i_y}.
\end{equation}
We note that the $\pi$ phase difference between the $c$ and $f$ electron transformation can be absorbed by transforming $V\to-V$. Consequently, each eigenenergy $E$ has a partner with a flipped sign $-E$. This is apparent from~\cref{eq:energy_equation}, since taking $\bm k \to \bm k+(\pi,\pi)$ flips the sign of even powers of $E$, such that $E\to-E$ is also an eigenenergy.

\textit{Reduction to the antiferrochannel Ansatz}---The above analysis reduces to the antiferrochannel case when setting $\bm{q}=(\pi,\pi)$. Explicitly, for $\bm{q}=(\pi,\pi)$, the dispersion obeys $\epsilon_{\bm k+ \bm q}=-\epsilon_{\bm k}$, which yields 
\begin{equation}
    E^3 - E \qty(\epsilon_{\bm k}^2 -V^2) = 0, 
\end{equation}
and the antiferrochannel bands are
\begin{equation}
    E^{1,2} = 0, E^{3,4} = -\sqrt{\epsilon_{\bm{k}}^2 + V^2}, E^{5,6} = +\sqrt{\epsilon_{\bm{k}}^2 + V^2}.
\end{equation}

\textit{Large-$N$ analysis of the order parameter}---To compute the channel symmetry-breaking order parameter in the large-$N$ limit, we introduce an external perturbation $J_a \to J_a + \Delta J_{a\bm{i}}$, where $\Delta J_{a\bm{i}}$ matches the $\pi$-modulated stripe channel. The corresponding order parameter can then be obtained from
\begin{equation}
    \expval{\mathcal{O}(\{a\bm{i}\})} =
    \frac{1}{\beta}\pdv{\log({Z})}{\Delta J_{a\bm{i}}}\eval{}_{\Delta J_{a\bm{i}}=0} = 
    \frac{\qty(V^*)^2}{J^2},
\end{equation}
with $V^*$ that minimizes $F(V)$.

\fi

\ifSM
\widetext
\newpage

\begin{center}
\textbf{\large Supplemental Materials: Antiferromagnetism and Stripe Channel Order in the $\SU(N)$-Symmetric Two-Channel Kondo Lattice Model}
\end{center}
\setcounter{secnumdepth}{1} 
\setcounter{equation}{0}
\setcounter{figure}{0}
\setcounter{table}{0}
\setcounter{section}{0}
\setcounter{page}{1}
\makeatletter
\renewcommand{\theequation}{S\arabic{equation}}
\renewcommand{\thefigure}{S\arabic{figure}}
\renewcommand{\bibnumfmt}[1]{[S#1]}
\renewcommand{\thesection}{S\arabic{section}}
\renewcommand{\citenumfont}[1]{#1}
\section{Next-nearest-neighbor hopping: large-$N$ analysis}

In this section, we examine the effect of next-nearest-neighbor (NNN) hopping of itinerant electrons on the stability of the $\pi$-stripe phase of the 2CKLM in the large-$N$ limit. To that end, we supplement the Hamiltonian with the term $H_{t'}$, defined as:
\begin{equation}
    H_{t'} = -t' \sum_{a=1}^2 \sum_{\mu}\sum_{\bm{i}}\sum_{\bm{j}\in\textrm{NNN}(\bm{i})}\qty(c^\dagger_{\bm{i}\mu a} c_{\bm{j}\mu a} +\text{h.c.}).
\end{equation}
Here, $\textrm{NNN}(\bm{i})$ denotes the diagonal neighbors on the square lattice. Specifically, for a site $\bm{i}=(i_x, i_y)$, this includes hopping events $(i_x \pm 1, i_y \pm 1)$. 

The generalized Hamiltonian does not respect particle-hole symmetry. Consequently, DQMC simulations are hampered by the numerical sign problem. For this reason, we proceed via a large-$N$ analysis to investigate the stability of the stripe phase.

We find that the $\pi$-modulated channel symmetry-breaking phase is stable for small hopping ratios, $t'/t \ll 1$. However, a sufficiently strong $t'/t$ eventually destabilizes the $\pi$-modulated phase in favor of a ferro-channel state. In~\cref{fig:J_tp}, we summarize our analysis. For each $t'/t$, we determine the critical coupling $J_c(t'/t)$ above which the $\pi$-modulated phase gives way to the ferro-channel state. The region in which the stripe phase survives shrinks as $t'/t$ increases, up to $t'/t=0.13$. Beyond this threshold, the ferro-channel phase dominates for all values of $J$ investigated ($J_c \to 0$).

\begin{figure}[b]
    \centering
    \includegraphics[width=0.5\linewidth]{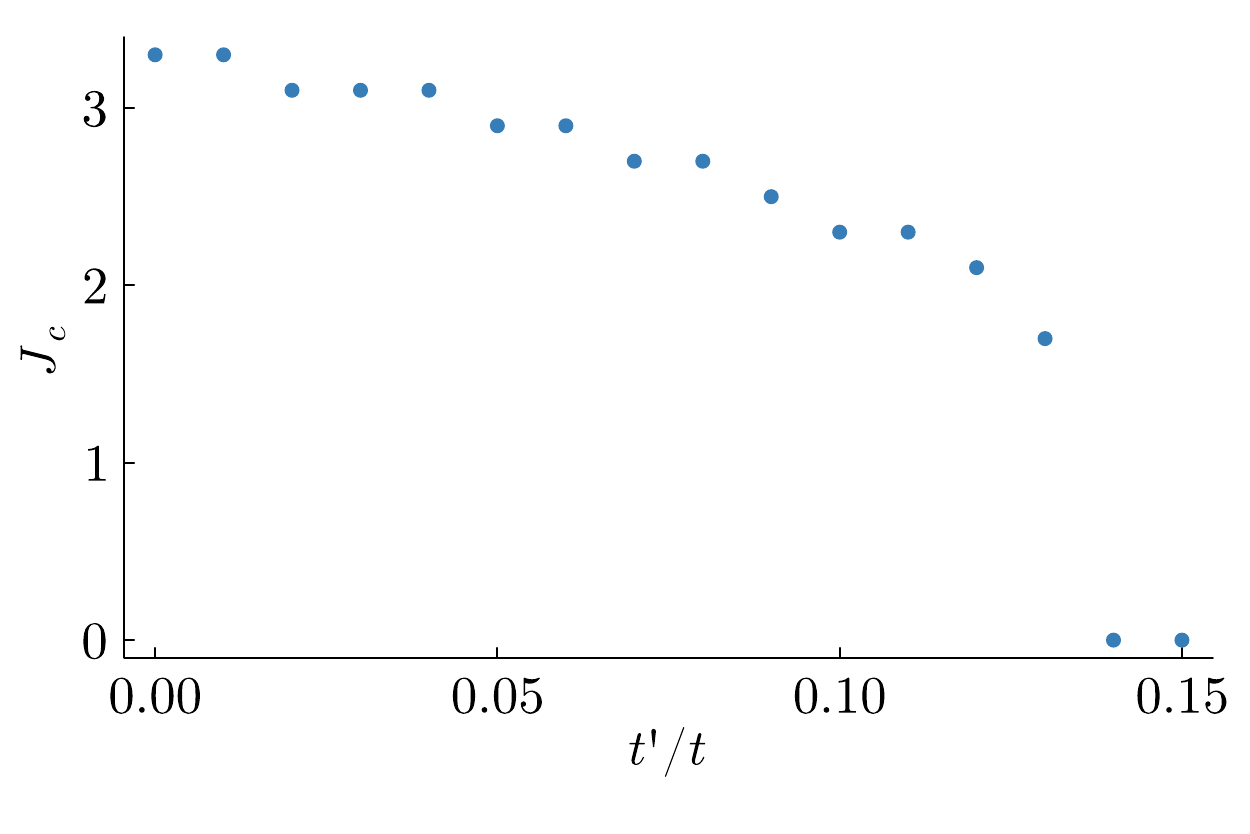}
    \caption{The critical value $J_c/t$, above (below) which, the ferrochannel (stripe) phase is preferred, is plotted as a function of $t'/t$.}
    \label{fig:J_tp}
\end{figure}

\section{Spatial structure of the channel symmetry breaking}

In the main text, we primarily focused on analyzing the dominant stripe order parameter associated with channel symmetry breaking. For completeness, in~\cref{fig:SM_dimer_all} we present the dimer correlation functions belonging to the three high‑symmetry wave vectors \(\bm{k}=(0,0)\), \((0,\pi)\), and \((\pi,\pi)\) for \(N=2,4,6\), and $8$. In all cases, we find that the only non-vanishing response is in the stripe order.

We also note an intriguing system-size-dependent evolution of the Bragg vector in the dimer correlations. While the stripe ordering wave vectors $\bm{k}=(0,\pi)$ and $(\pi,0)$ dominate for system sizes $L=6,10,14,$ and $18$, we find that for $L=4,8,12,$ and $16$, the maximum in the dimer correlation function, shifts to the nearest available momenta around these stripe wave vectors. Importantly, both sequences converge in the thermodynamic limit to the same $(0,\pi)$ stripe ordering.

To accelerate convergence, we exclude system sizes $L=4,8,12,$ and $16$, relying solely on $L=6,10,14$, and $18$. In~\cref{fig:even_odd}, we present results for various system sizes in the stripe phase. In this dimension and for half-filled stripes, lattice sizes $L = 4n+2$ correspond to filled shell configurations and are known to provide a nice sequence of sizes to extrapolate to the thermodynamic limit.

\begin{figure}[b]
    \centering
    \includegraphics[width=\linewidth]{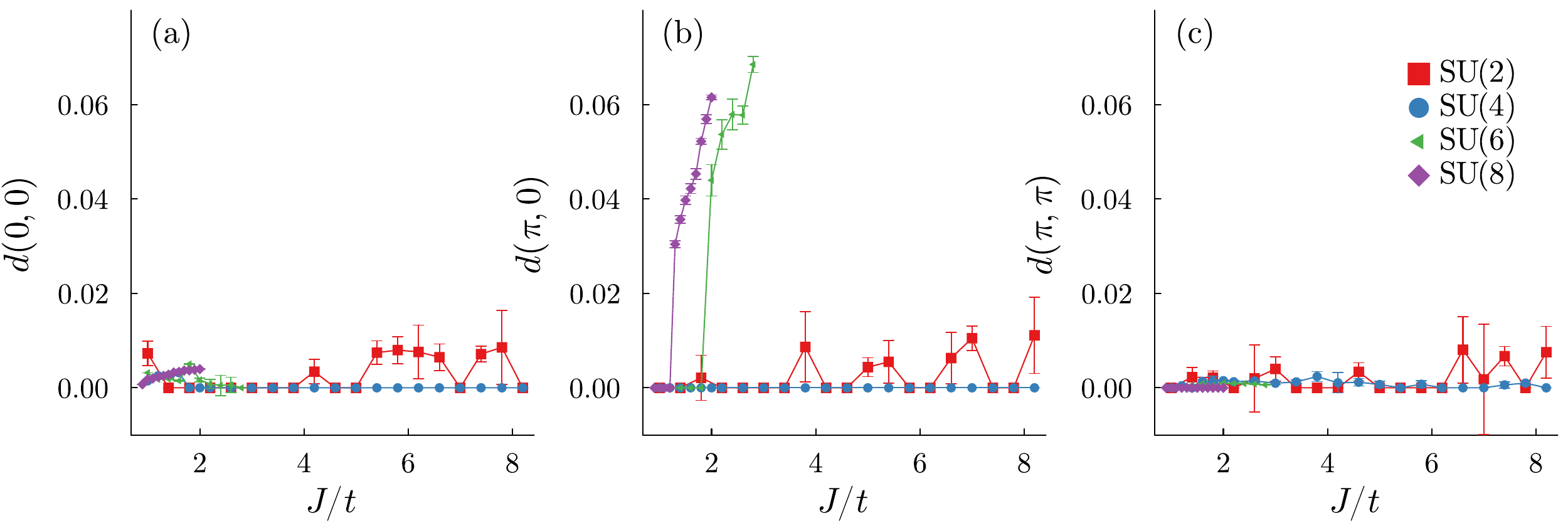}
    \caption{Dimer order parameter function at momentum 
    $\bm{k}=(0,0)$~(a), 
    $(0,\pi)$~(b), 
    and $(\pi,\pi)$~(c) as a function of $J/t$. Different curves correspond to $\SU(N)$ symmetric model with $N=2,4,6$, and $8$.}
    \label{fig:SM_dimer_all}
\end{figure}

\begin{figure}
    \captionsetup[subfigure]{labelformat=empty}
    \subfloat[\label{subfig:even_odd_dimer_1d}]{}
    \subfloat[\label{subfig:even_odd_dimer_heat}]{}
    \centering
    \includegraphics[width=0.48\linewidth]{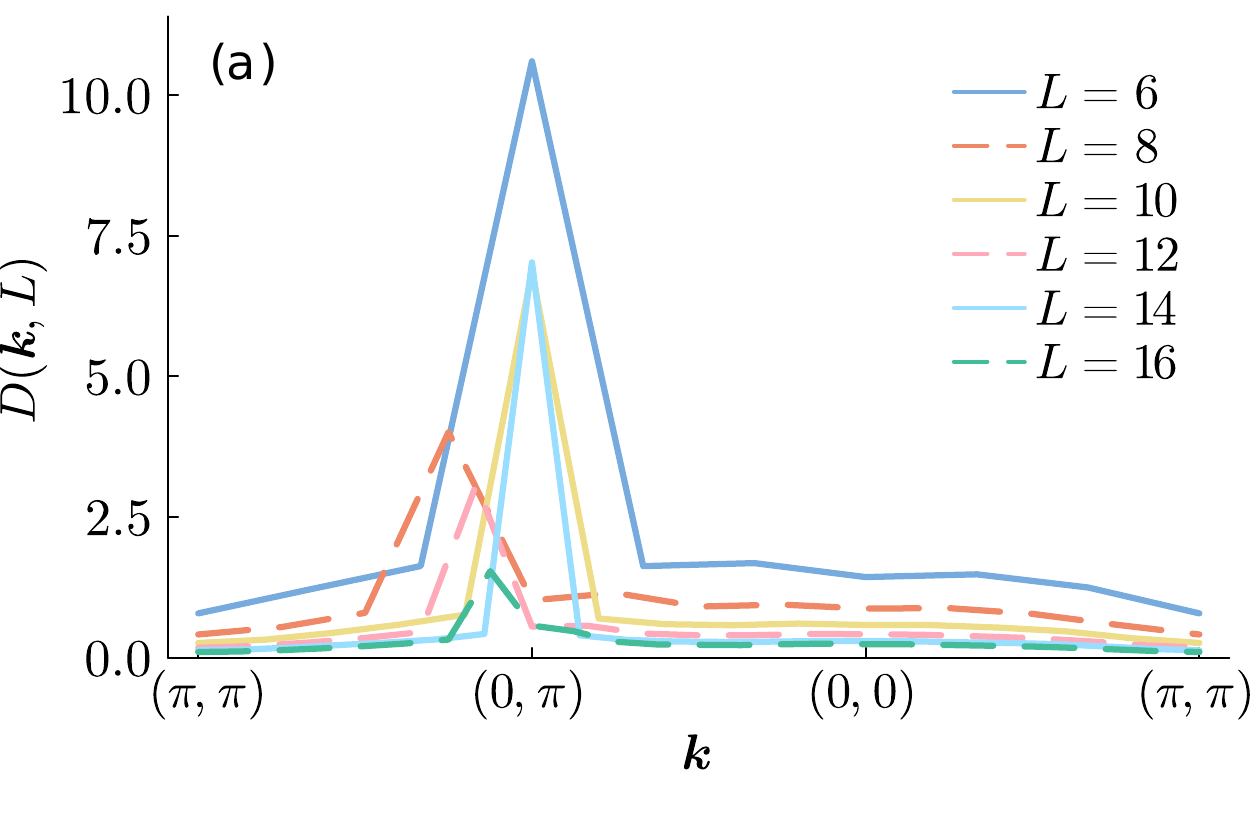}
    \includegraphics[width=0.48\linewidth]{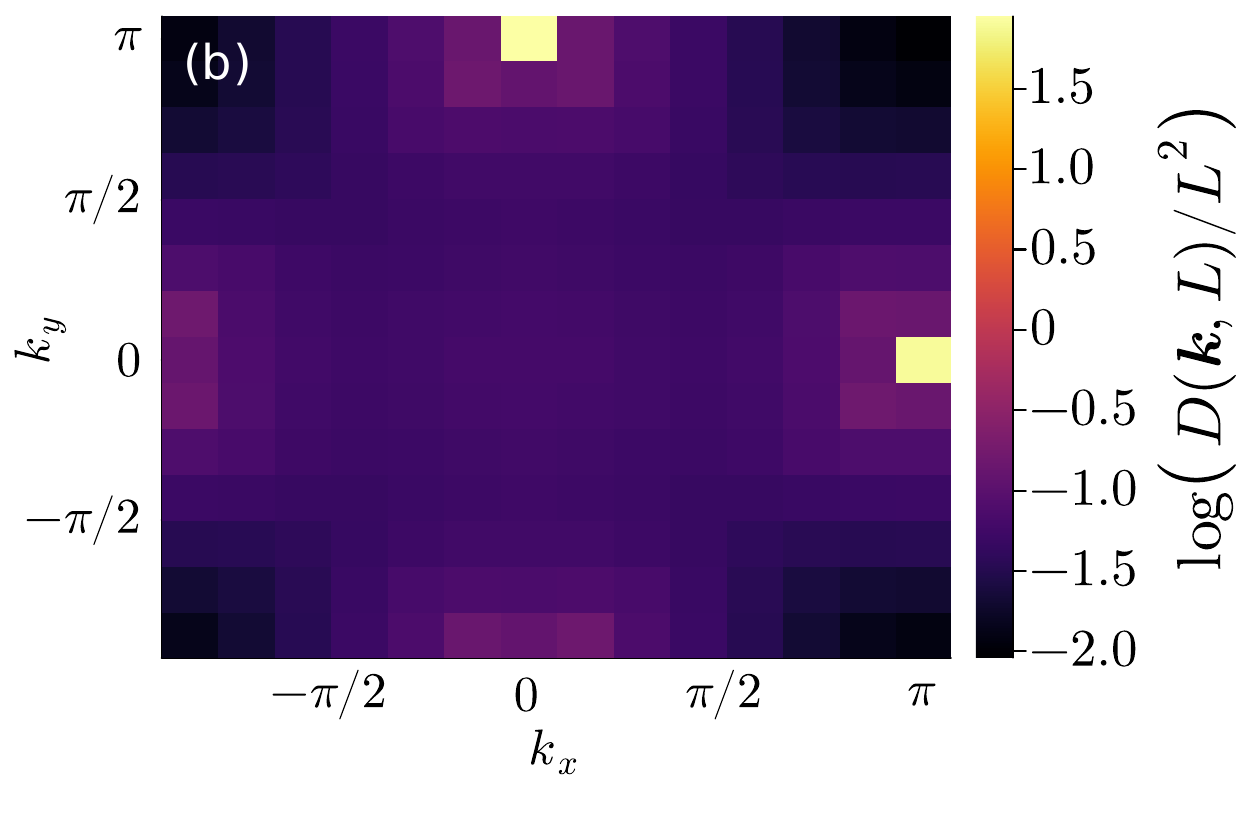}

    \caption{(a) Dimer–dimer correlation function plotted along a high-symmetry path in momentum space, different curves correspond to varying system sizes. In the thermodynamic limit, the Bragg peak approaches the wave vector with momentum $\bm{k}=(0,\pi)$. Simulations are performed in the stripe phase, at coupling strength $J/t=1.6$ larger than $J^{\textrm{stripe}}_c/t = 1.3$ of the $\SU(8)$ symmetric model. Solid lines correspond to system sizes $L = 4n+2$ sequence, for integer $n$, which we used in our extrapolation to the thermodynamic limit.
    (b) Heatmap of the logarithm of the dimer correlation function in $k$-space for $L=14$, showing pronounced peaks at $\bm{k}=(0,\pi)$ and $\bm{k}=(\pi,0)$, consistent with stripe ordering.}
    \label{fig:even_odd}
\end{figure}
\section{Dimer correlations of $\SU(4)$}
As detailed in the main text, we did not observe long-range dimer correlations for $N\leq4$. Nonetheless, as shown in ~\cref{fig:su4_dimer}, we find that the strongest response in $d(\bm{k})$ is at $\bm{k}=(\pi,0)$ followed closely by $\bm{k}=(0,0)$. This observation can serve as a precursor of the $\pi$-modulated stripe ordering for $N=4$.
\begin{figure}
    \centering
    \includegraphics[width=0.5\linewidth]{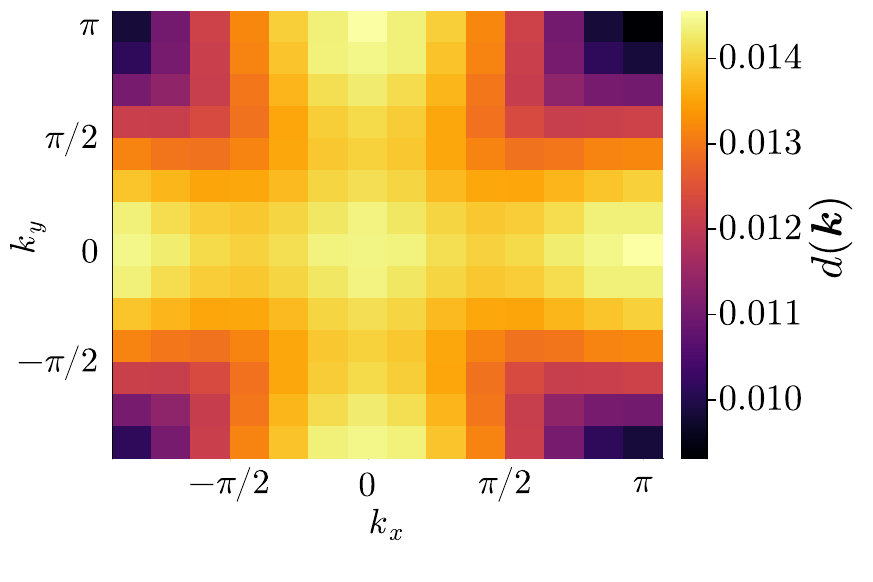}
    \caption{Heatmap of the dimer order parameter in ${k}$-space for $\SU(4)$, $L=14$, and $J/t = 1.6$.}
    \label{fig:su4_dimer}
\end{figure}
\section[]{Additional details of the Monte Carlo simulation and finite-size scaling analysis}

\titleformat{\subsection}[runin]
  {\normalfont\bfseries}
  {}
  {0.5em}
  {}
  [ --]
\titlespacing{\subsection}{0pt}{1.5ex plus 0.2ex}{0.5em}

In this section, we provide additional technical details pertinent to the DQMC simulation and data analysis presented in the main text.
\subsection[]{Extrapolation to the thermodynamic limit}

From our finite‑size scaling analysis, we extract the values of various observables in the thermodynamic limit using the standard \textit{Ansatz}
\begin{equation}
    O(L) = \mathcal{O}_\infty + \frac{A_1}{L^2}+\frac{A_2}{L^4}.
    \label{eq:L_fit}
\end{equation}
Here $O(L)$ and $\mathcal{O}_\infty$ represent the finite-size and thermodynamic values, respectively. For non-vanishing observables, $\mathcal{O}_\infty>0$, we find that omitting the smallest system size $L\leq6$ improves the fit quality. To avoid overfitting, in some cases, we omit higher-order corrections by setting $A_2=0$. We showcase this procedure in~\cref {fig:largeL}. 

\begin{figure}
        \centering{
        \includegraphics[width=0.9\linewidth]{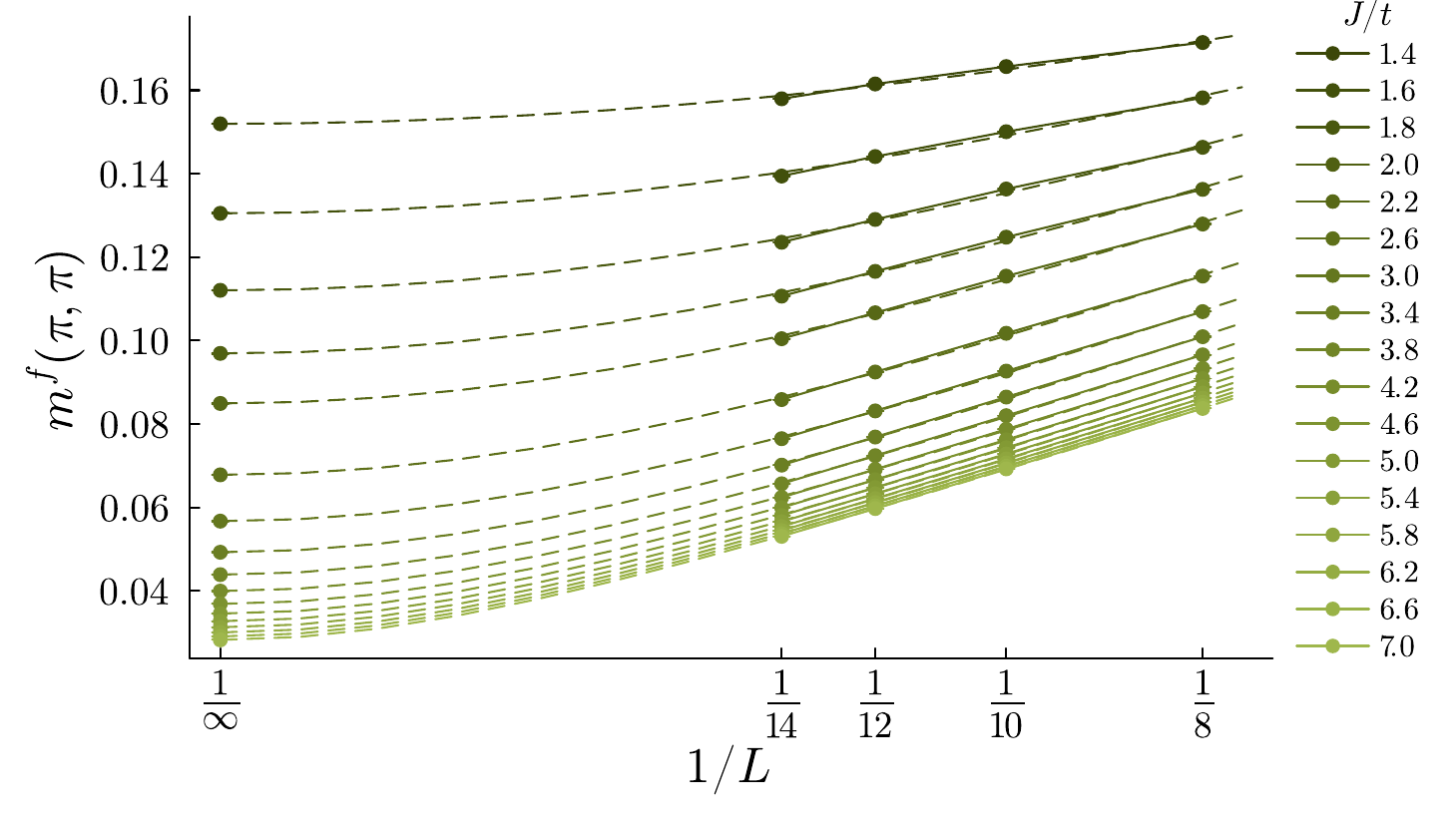}}
        \caption{Finite-size scaling of the spin-spin correlation function at momentum $\bm{k}=(\pi,\pi)$ as a function of the inverse linear system size $1/L$. Different curves correspond to different Kondo coupling values $J/t$. Dashed lines represent a quartic fit, see~\cref{eq:L_fit}.}
        \label{fig:largeL}
\end{figure}

\subsection[]{Comparison with exact diagonalization in the single site limit}

To benchmark and validate our implementation, we study the single-site limit ($t=0$). In this limit, we can compute observables exactly by diagonalizing the local Hamiltonian and comparing them directly with DQMC results.

We focus on the spin and dimer correlation functions. The on-site spin-spin correlation function between the channels and the localized moment is defined similarly 
\ifmainText
to~\cref{eq:ss_corr}
\else
in~Eq.~(4)
\fi
in the main text, but for a single site. For the dimer-dimer correlation function, we consider the components of ${M_z}$ defined 
\ifmainText
in~\cref{eq:M_vec}
\else
in~Eq.~(5)
\fi
in the main text. Explicitly, 
\begin{equation}
    M_{z} = 
    \mathcal{D}_1
    -
    \mathcal{D}_2,
\end{equation}
where 
\begin{equation}
    \mathcal{D}_a = 
    c_{\mu a}^\dagger
    c_{\mu'a} 
    ({T}_{\mu\mu'}^\sigma
    {T}_{\nu\nu'}^\sigma 
    )
    f^\dagger_{\nu}
    f_{\nu'}.
\end{equation}

We examine the following quantity,
\begin{equation}
    \left \langle \mathcal{D}_a\mathcal{D}_b \right \rangle = 
    \sum_{\alpha,\beta,\gamma,\delta} 
    \left \langle 
    S^\alpha_{~\beta}(c_a)
    S^\beta_{~\alpha}(f)
    S^\gamma_{~\delta}(f)
    S^\delta_{~\gamma}(c_b)
    \right \rangle, 
\end{equation}
where $a,b=1,2$ denote the channel indices. Here, $S^\alpha_{~\beta}(\psi)$ are the $\SU(N)$ generators as 
\ifmainText
in~\cref{eq:constraint}
\else
in~Eq.~(2)
\fi
of the main text. The quantity $\left \langle \mathcal{D}_a\mathcal{D}_b \right \rangle$ captures the correlation between dimers formed in channels $a$ and $b$.

In~\cref{fig:t0_spin} and~\cref{fig:t0_dimer}, we compare the spin-spin and dimer-dimer correlations computed from DQMC simulations to the exact diagonalization results for $N=2$ and $N=4$, plotted as a function of the ratio $J_1/J_2$. In all cases, we observe excellent agreement, validating our DQMC implementation at the single-site level.

\begin{figure}
    \centering
    \includegraphics[width=0.495\linewidth]{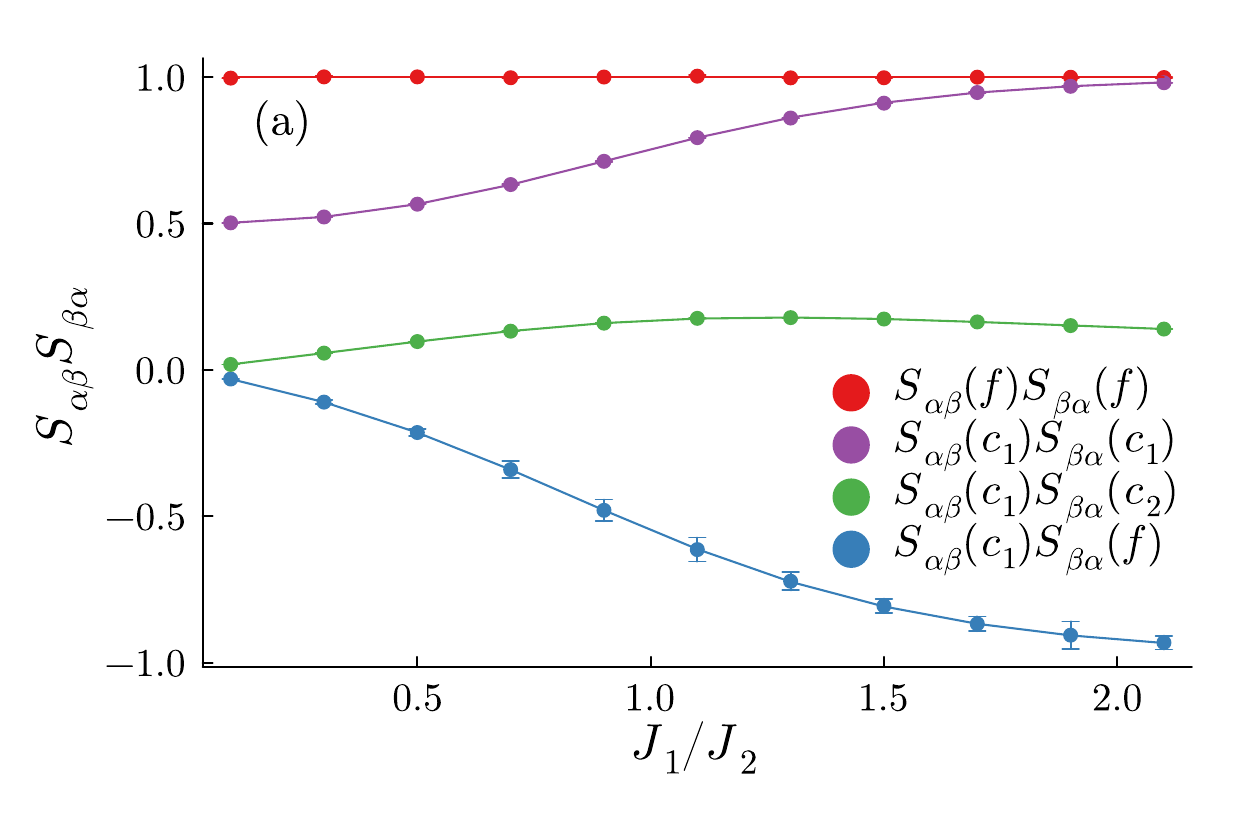}
    \includegraphics[width=0.495\linewidth]{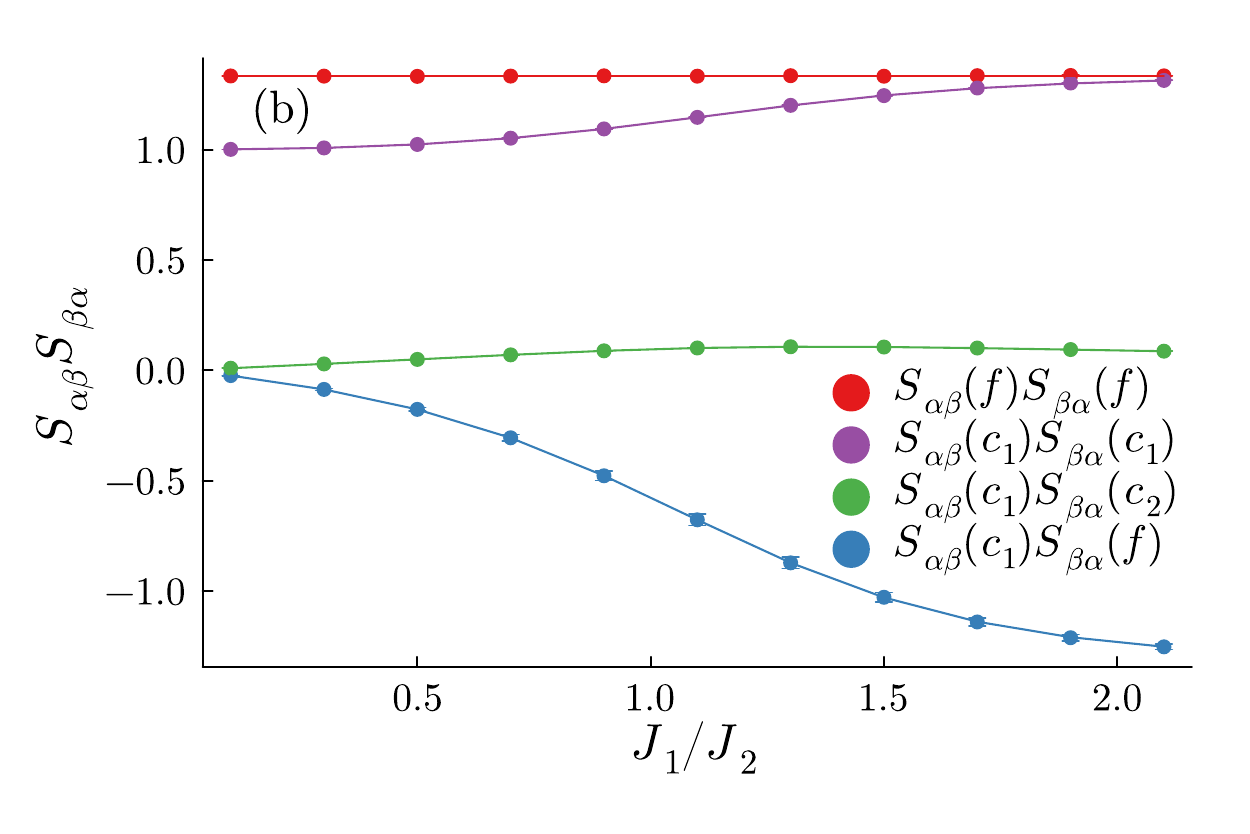}
    \caption{Spin-spin correlation function between the f- and c-electrons as a function of Kondo coupling $J_1$, with $J_2=1$. Results are shown for (a) $N=2$ and (b) $N=4$. Different curves correspond to different cross-correlations between the channels. Here, $J_2 \beta=4$ and $U_f/J_2=10$. Dots: DQMC results; and solid lines: exact diagonalization.}
    \label{fig:t0_spin}
\end{figure}
\begin{figure}
    \centering
    \includegraphics[width=0.495\linewidth]{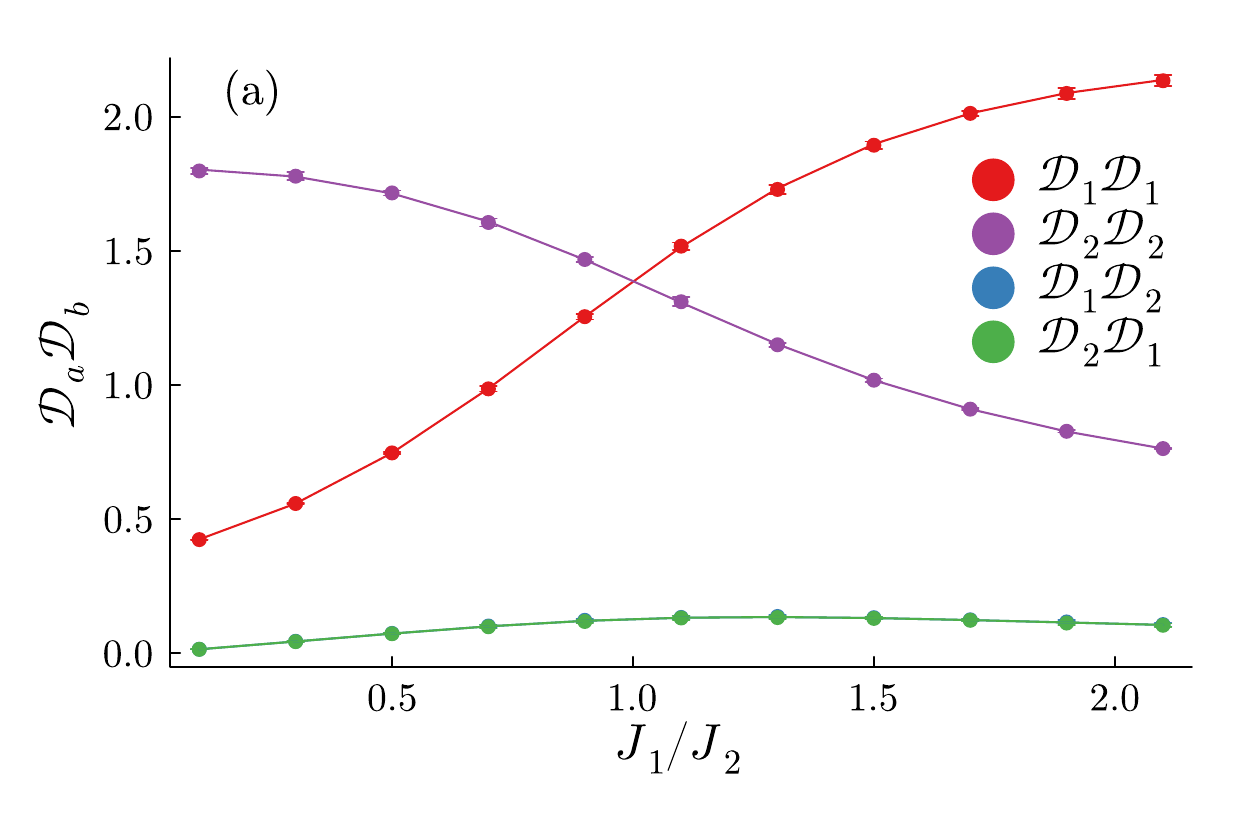}
    \includegraphics[width=0.495\linewidth]{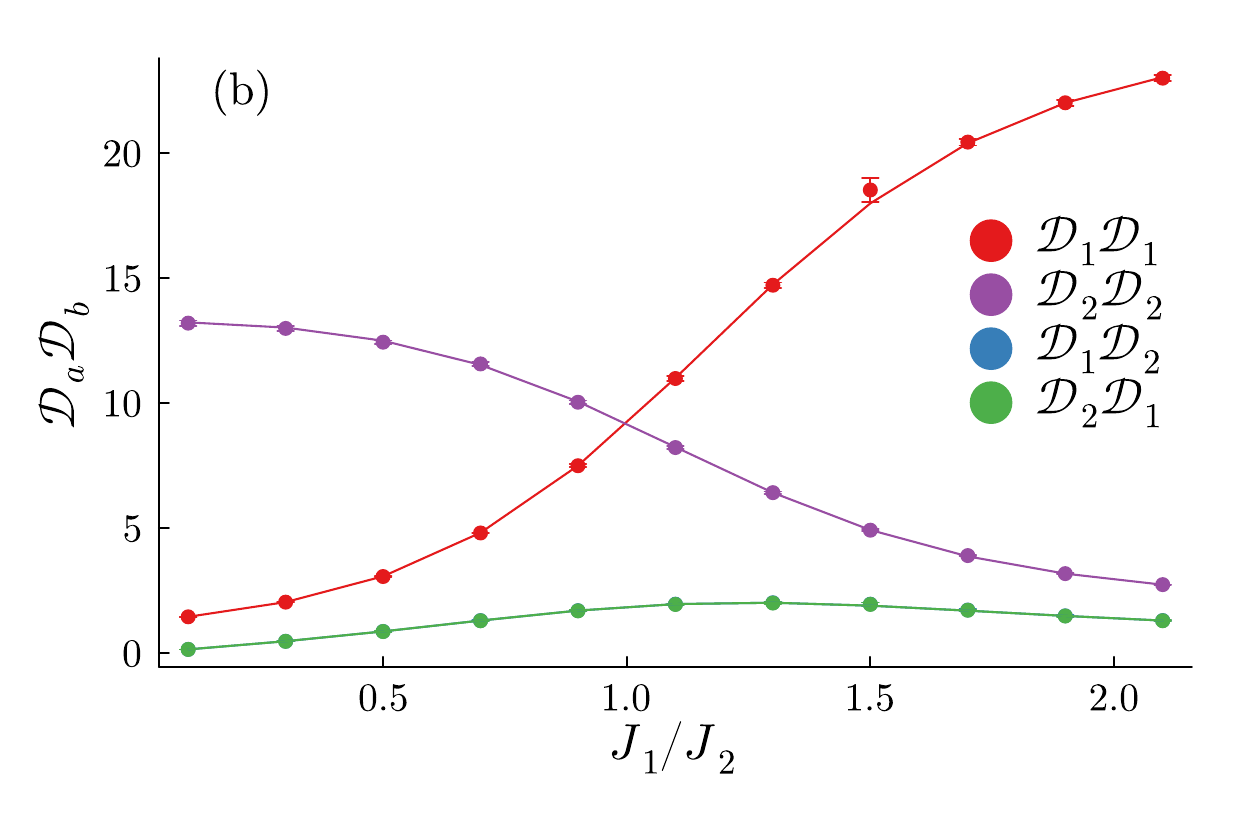}
    \caption{Dimer-dimer correlation function between the f- and c-electrons as a function of Kondo coupling $J_1$, with $J_2=1$. Results are shown for (a) $N=2$ and (b) $N=4$. Different curves correspond to different cross-correlations between the channels. Here, $J_2 \beta=4$ and $U_f/J_2=10$. Dots: DQMC results; and solid lines: exact diagonalization.}
    \label{fig:t0_dimer}
\end{figure}

\subsection[]{Large-\texorpdfstring{$U$}{U} limit}

To ensure the validity of the half-filling constraint on the localized $f$-electrons, we performed simulations for several values of the on-site repulsion $U_f$. In~\cref{fig:SM_constraint_N28_U}, we show the variance of the local $f$-electron occupancy,
\begin{equation}
    \frac{1}{L^2}\biggl\langle \sum_{\bm{i}} \Bigl(n^f_{\bm{i}} - \frac{N}{2}\Bigr)^2 \biggr\rangle,
\end{equation}
as a function of $J/t$ for $N=2$ and $N=8$. Our results show negligible occupancy of the $f$-electrons for $L=6$, and that $U_f=4$ used in our simulation is sufficient for convergence. Additionally, we verify that increasing the system size to $L=14$ (dashed line in the right panel) results in even lower occupancy.
\begin{figure}
    \centering
    \includegraphics[width=0.8\linewidth]{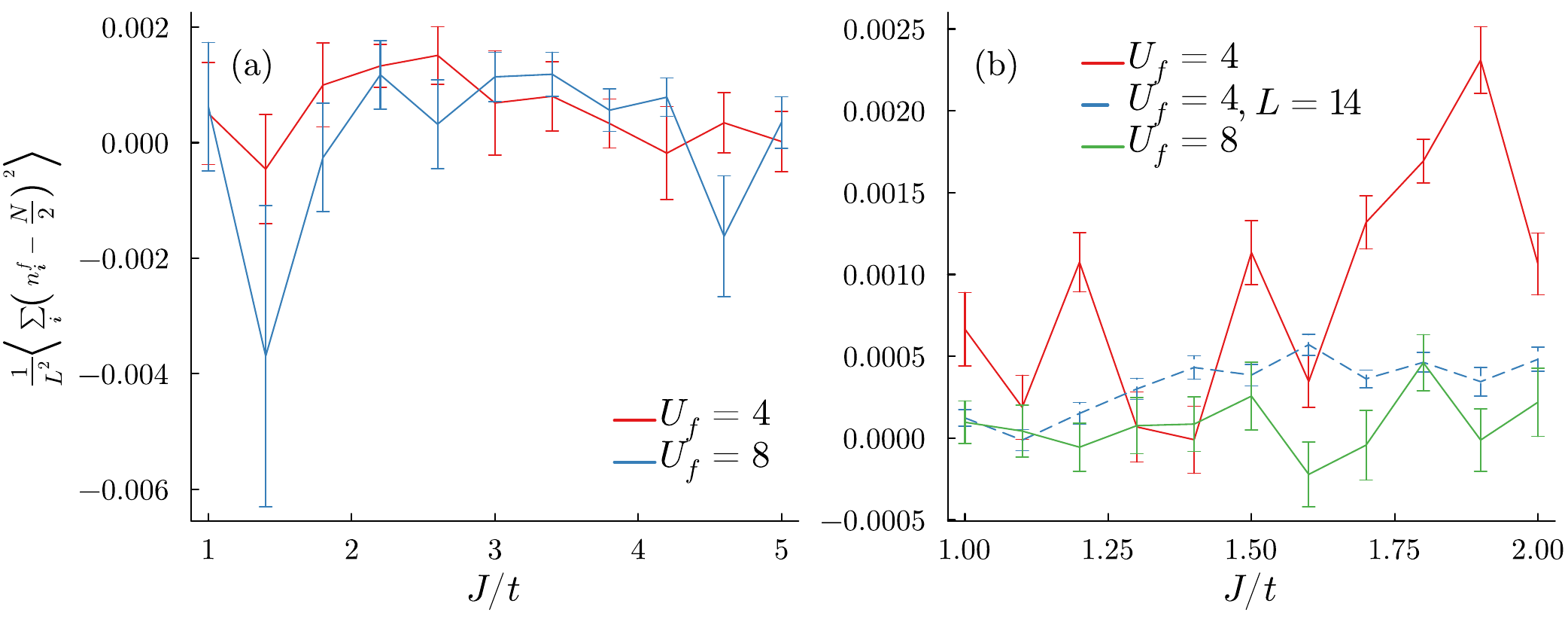}
    \caption{Variance of the local $f$-electron occupancy, as a function of $J/t$ for $N=2$ (a) and $N=8$ (b). Results are shown for different values of $U_f$. Here, most simulations use system size $L=6$; the dashed line in the right panel shows data for $L=14$ at $U_f=4$.}
    \label{fig:SM_constraint_N28_U}
\end{figure}

\begin{figure}
    \centering
    \includegraphics[width=1\linewidth]{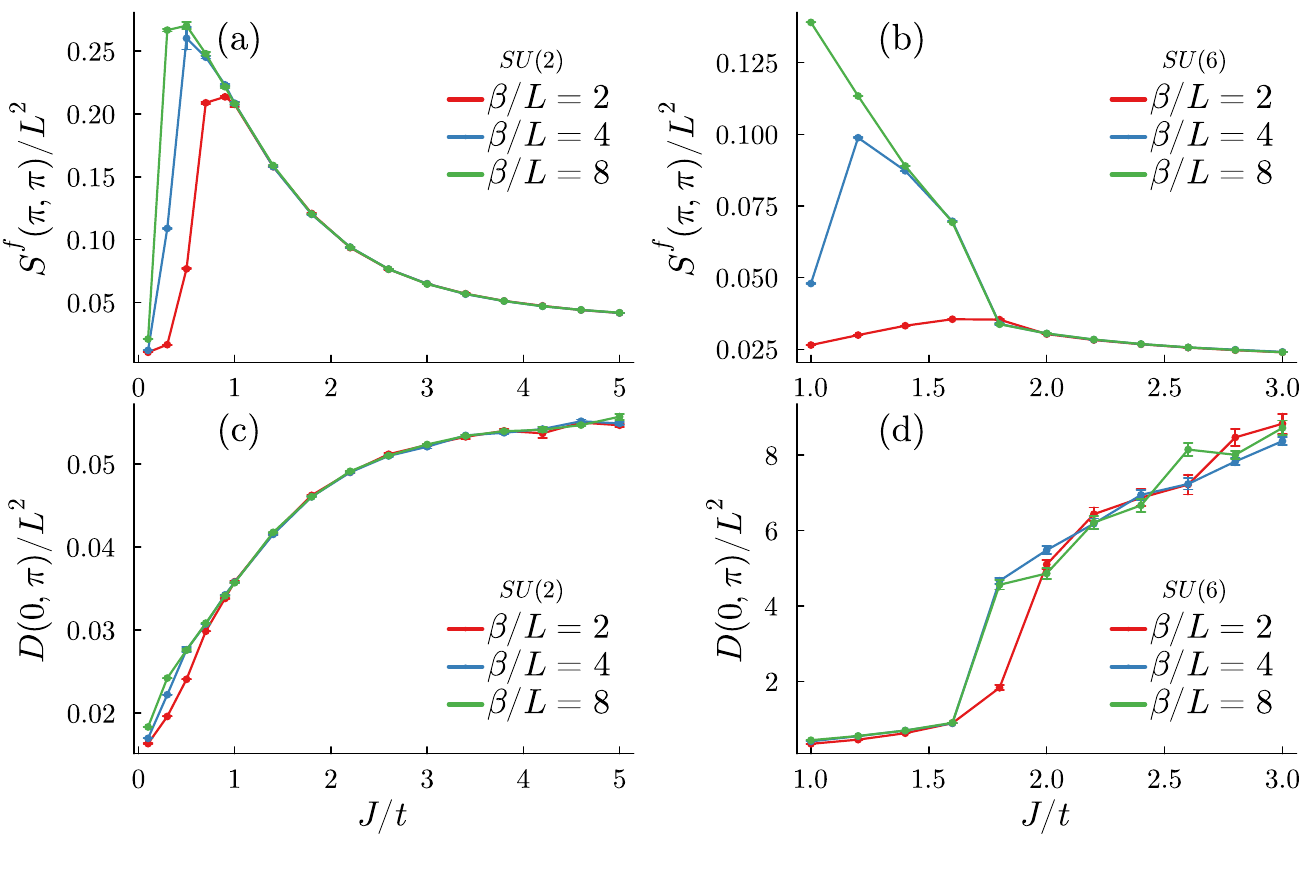}
    \caption{Convergence test to the zero temperature limit. We examine the spin-spin correlation function at momentum $(\pi,\pi)$ (a) and the dimer-dimer order parameters at momentum $(0,\pi)$ (c) for the $\SU(2)$ case and for the $\SU(6)$ case in (b) and (d). At low $J/t$, lower temperatures are required to capture the AFM phase accurately. Different curves correspond to different $\beta/L$ ratios. All results correspond to a system size of $L=10$.}
    \label{fig:SM_SUN_beta}
\end{figure}
\subsection[]{Trotter error}
As is standard, we employ a Trotter decomposition by discretizing the imaginary‐time interval \(\beta\) into \(M\) slices of width \(\varepsilon\), such that
$\beta=M\varepsilon$.
Throughout our simulations, we chose
\begin{equation}
  \varepsilon =\min\qty(0.1,0.1/J),
\end{equation}
which reduces \(\varepsilon\) at larger coupling \(J/t\). Convergence tests with \(\varepsilon/2\) produce results that agree within the statistical error bars, confirming that the effect of Trotter errors is negligible compared to statistical errors.

\subsection[]{Low temperature limit}

Throughout the simulations, the inverse temperature $\beta=1/T$ was chosen to be sufficiently large to reproduce ground-state properties within the error bars. We find that taking $t\beta/L=4$ was adequate for the majority of the parameter space.
However, the low-coupling regime necessitates larger $\beta$; we adopted $t\beta/L=8$ for cases where $J/t \leq 1.5$ and $N=8$.
In~\cref{fig:SM_SUN_beta}, we present results for different $\beta/L$ ratios for the $\SU(2)$ and $\SU(6)$ case.

\subsection[]{Statistical analysis}
In our simulations, we discarded the initial transient period to ensure thermalization. Following this equilibration phase, we collected between 1,700 and 3,000 Monte Carlo measurements. Statistical uncertainties were then estimated using a binning and bootstrap analysis. 

\fi
\end{document}